\documentclass[preprint,12pt]{elsarticle}
\usepackage{amssymb,latexsym,amsmath}
\usepackage{theorem}
\usepackage{empheq}
\usepackage{ifthen}
\usepackage{amsfonts}
\usepackage{latexsym}
\usepackage{makeidx}
\usepackage{graphicx}
\usepackage{epsfig,float}
\newcommand{\D}{\Delta}
\newcommand{\Om}{\Omega}
\newcommand{\p}{\phi}
\renewcommand{\a}{\alpha}

\renewcommand{\t}{\tau}

\newcommand{\om}{\omega}

\newcommand{\e}{\varepsilon}

\begin{document}
%%%%%%%%%%%%%%%%%%%%%%%%%%%%%%%%%%%%%%%%%%%%%%%%%%%%%%%%%%%%%%%%%%%%%%%%%%%

\begin{frontmatter}

\title{Ant foraging and minimal paths in simple graphs}

\author[urjc]{M.~Vela-P\'{e}rez\corref{cor1}}
\ead{mvp$\_$es@yahoo.es, mvela@profesor.ie.edu}
\author[csic]{M.~A.~Fontelos}
\ead{marco.fontelos@icmat.es}
\author[csic]{J.~J.~L.~Vel\'{a}zquez}
\ead{jj$\_$velazquez@icmat.es}

\cortext[cor1]{Corresponding author}

\address[urjc]{Departamento de Arquitectura, IE University, C/ Z\'{u}\~{n}iga 12, 40003 Segovia, Spain}
\address[csic]{Instituto de Ciencias Matem\'{a}ticas, (ICMAT, CSIC-UAM-UC3M-UCM), C/ Nicol\'{a}s Cabrera 15, 28049 Madrid, Spain}

\date{\today}

%-----------------------------------------------------------------------
\begin{abstract}
Ants are known to be able to find paths of minimal length between the nest and food sources.
The deposit of pheromones while they search for food and their chemotactical response to them
has been proposed as a crucial element in the mechanism for finding minimal paths.
We investigate both individual and collective behavior of ants in some simple networks representing basic mazes. The character of the graphs considered is such that it allows a fully rigorous mathematical treatment via analysis of some markovian processes in terms of which the evolution can be represented.
Our analytical and computational results show that in order for the ants
to follow shortest paths between nest and food, it is necessary to superimpose to the ants' random walk the
chemotactic reinforcement. It is also needed a certain degree of persistence so that ants tend to move preferably without
changing their direction much. It is also important the number of ants, since we will show that the speed for finding minimal paths increases very fast with it.
\end{abstract}
%-----------------------------------------------------------------------

%-----------------------------------------------------------------------
%-----------------------------------------------------------------------

\end{frontmatter}

%%%%%%%%%%%%%%%%%%%%%%%%%%%%%%%%%%%%%%%%%%%%%%%%%%%%%%%%%%%%%%%%%%%%%%%
\section{Introduction}
%%%%%%%%%%%%%%%%%%%%%%%%%%%%%%%%%%%%%%%%%%%%%%%%%%%%%%%%%%%%%%%%%%%%%%%

Transport networks play an important role in different natural and man-made systems. In the last years many work has been done to understand collective patterns generated by the individual workers' trail laying, showing how complex collective structures in insect colonies may be based on self-organization and co-operation \cite{GADP}. Foraging ants find the shortest paths for initially unknown food sources in almost the minimum possible time for certain types of mazes (\cite{GGCFT} and \cite{GADP}). How can an animal with only limited and local information achieve this in such an efficient way? Many ants, having only a limited individual capacity for orientation are able to select the shortest path between nest and food source dodging many obstacles by just following the pheromone trail. Just as the functioning and success of modern cities are dependent on an efficient transportation system, the effective management of traffic is also essential to ant colonies.

There are different types of ants that behave in a different way. In the last years, many experimental results have been developed related to, among others, Argentine ant (\emph{Iridomyrmex humilis}) \cite{APD,DAGP,GGCFT,VTGFAT}, Pharaoh's ant (\emph{Monomorium pharaonis}) \cite{JRD,RRH}, Lasius niger (\emph{Hymenoptera,Formicidae}) \cite{BDG,DFHD,ND} and Army ant (\emph{Eciton burchelli}) \cite{B,CF,FGGD}.

It has been proved that different ants use one pheromone (Argentine ant, Lasius niger and Army ant) whereas others employ three types of pheromone (Pharaoh's). This pheromone has a mean lifetime larger compared to the time spent for the ants to move from nest to food source and so ants
can reinforce the geodesic path.

In \cite{GADP}, a series of experiments have been done with Argentine ant in special mazes consisting of graphs.
As it is well known, Argentine ant has a limited individual capacity for orientation. Hence, they need to cooperate via pheromone trails with other ants in order to find the shortest path to the food source. The authors posed a model consisting on a system of ordinary differential equations for a graph, which is derived as a mean field theory of a stochastic model and it is solved numerically.

There are not many studies concerning motion of ants in the plane. They are mostly concerned with the particular case of the Army ant. These colonies of ants are huge (may have a million of workers) and carnivores, and form traffic lanes in their main foraging columns. In \cite{CF} it is shown that the movement rules of individual ants can produce a collective behavior creating distinct traffic lanes that minimize congestion and maximize traffic flow. This is done assuming pre-existing pheromone concentration with fixed profile. A general model of ant behavior is developed, in terms of individual-based simulation approach. To do so, it is studied first the behavior of individual ants in the absence of interactions with other ants. After so, the collective properties of the model during the generation of spatial patterns are investigated.
In this model it is shown how local interactions and individual movement rules can strongly influence the organization of traffic over a large spatial scale. Nevertheless, it does not constitute a complete model due to the fact that pheromone concentration is assumed to be fixed in time and
the formation of such concentration is not explained.

All these observations pose the mathematical problem of determining a minimal set of rules so that a given number of ants following them tend to choose shortest paths between nest and food source. From the experimental observations it seems that such mechanism should include the presence of pheromone and the persistence (tendency to follow straight paths in the absence of other effects).
Remarkably this effect has been invoked in the past to explain the formation of filamentary structures in some biological problems such as the formation of vascular networks \cite{SAGGPB}.

We will consider ants as random walkers where the probability to move in one or another direction is influenced by the concentration of pheromone near them. This kind of motion is known in the mathematical literature as \emph{reinforced random walks}. There is a vast amount of work in this area (see for instance the review \cite{P}, the seminal paper \cite{Da} or \cite{V} for random walks in graphs). The direct relation of reinforced random walks with biology was stressed in \cite{OS}, where general rules were found for obtaining chemotactical aggregation in a single point. In our study, we are mainly interested not in an individual random walker but rather on a large number of random walkers, their collective behavior, and the possibility for them to aggregate forming geodesic paths between two points. Our work relates to current research on swarming, flocking and general motions of brownian agents but with essential differences derived from the fact that it is chemical signals (instead of visual, acoustic, or other type), coupled with a directional bias in the random walk process, what tends to produce paths of minimal length.

The purpose of this article is to show rigorously how the combined effect of reinforced random walks and persistence is able to produce
the selection of paths of minimal length in simple networks. To do so we investigate the behavior of ants in a two node network and in a three node network (with and without directionality constraint). The paper is organized as follows: in Section $2$ we will do some numerical experiments for a two node network and a three node network to understand the role of each parameter of the model. In Sections $3$ and $4$ we prove some analytical results to find the minimum number of ingredients that are required to obtain preference for the shortest paths. Section $3$ is devoted to finding the possible long-time dynamics while section $4$ is concerned with the dynamics at early times. Finally, in Section $5$ we summarize our work and point to future directions which might be of interest.

%%%%%%%%%%%%%%%%%%%%%%%%%%%%%%%%%%%%%%%%%%%%%%%%%%%%%%%%%%%%%%%%%%%%%%%
\section{Numerical results}
%%%%%%%%%%%%%%%%%%%%%%%%%%%%%%%%%%%%%%%%%%%%%%%%%%%%%%%%%%%%%%%%%%%%%%%

We study numerically the collective behavior of a variable number of ants in networks in the form of graphs with $E$ edges, $w_1,w_2,\ldots,w_E$. The experiments are done using a Monte Carlo method with the random number generator \emph{binornd} from MATLAB. This random number generator returns numbers from a binomial distribution with parameters $N$ (number of Yes/No experiments) and $p$ (probability of success). We perform a certain number of simulations for a given number of time steps and for a number $H$ of ants.

Our purpose is to explain the behavior of ants choosing the shortest path in terms of reinforcement and directionality constraints. We explore in detail what is the role of each parameter and how they affect to the collective behavior.

%%%%%%%%%%%%%%%%%%%%%%%%%%%%%%%%%%%%%%%%%%%%%%%%%%%%%%%%%%%%%%%%%%%%%%%
\subsection{Simulations for a two node network}
\begin{figure} [h!]
\centering
    \includegraphics[width=0.5\textwidth]{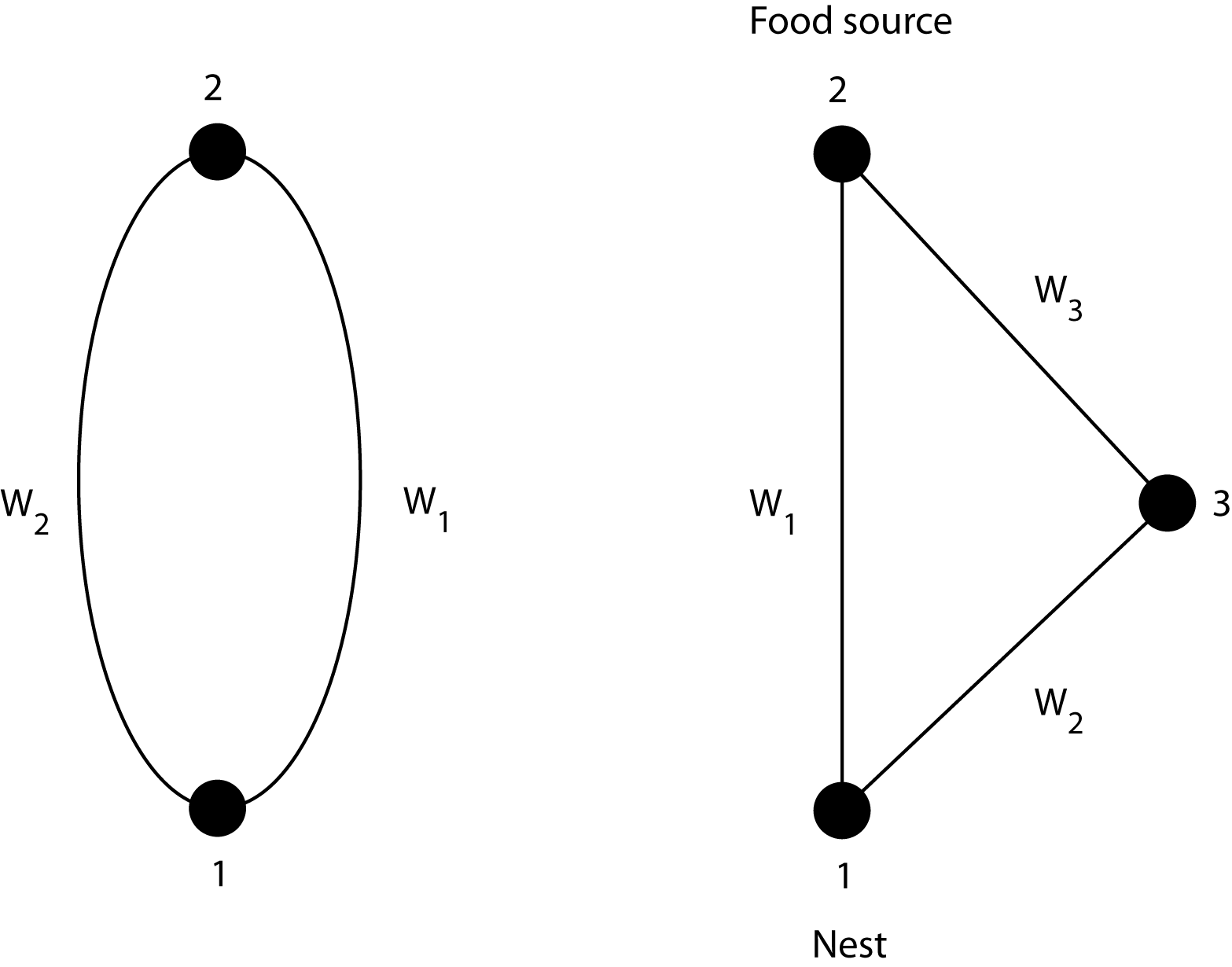}
    \caption{Two node network (left) and three node network (right)}
    \label{network}
\end{figure}

%%%%%%%%%%%%%%%%%%%%%%%%%%%%%%%%%%%%%%%%%%%%%%%%%%%%%%%%%%%%%%%%%%%%%%%
We consider a two node network with reinforcement as in figure \ref{network} left.
Let $p_{W_1}$ be the probability of moving from node $1$ to node $2$ through the edge $W_1$ and $p_{W_2}$ be the probability of moving from node $1$ to node $2$ through the edge $W_2$.

Following \cite{DAGP} and \cite{GGCFT}, we take the probabilities at step $t$ to be
\begin{equation}\label{prob:red2:eq1}
p_{W_1}(t)=\frac{(k+\om_1(t))^\a}{(k+\om_1(t))^\a+(k+\om_2(t))^\a},
\end{equation}
\begin{equation}\label{prob:red2:eq2}
p_{W_2}(t)=\frac{(k+\om_2(t))^\a}{(k+\om_1(t))^\a+(k+\om_2(t))^\a},
\end{equation}
where $\om_1(t),\om_2(t)$ are the quantities of pheromone at each link $W_1$, $W_2$ respectively at time $t$, $k$ is a positive constant and $\a$ is the exponent of the non-linearity. The value of $\om_1(t)$ (resp. $\om_2(t)$) is increased in one unit each time the ant moves along the edge $W_1$ (resp. $W_2$), representing the deposit of pheromone by the ant.

%---------------- INSERTAR FIGURA DE LA RED ---------------------------%
%---------------------------------------------------------------------%

\begin{figure} [h!]
\centering
    \includegraphics[width=0.7\textwidth]{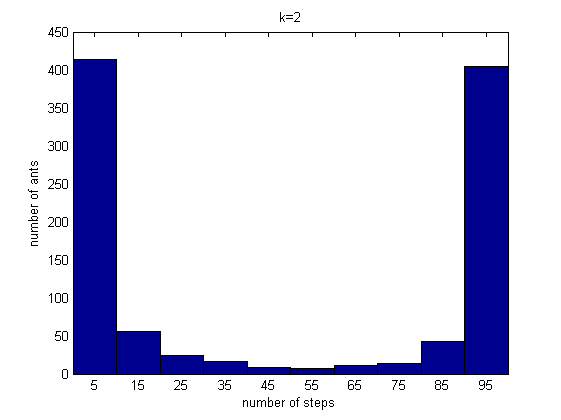}
\caption{Histogram for $k=2$, $\a=2$, $n=100$ time steps. The experiment is repeated $1000$ times and we represent the cumulative result.}
 \label{simu1}
 \end{figure}

\begin{figure} [h!]
\centering
    \includegraphics[width=0.75\textwidth]{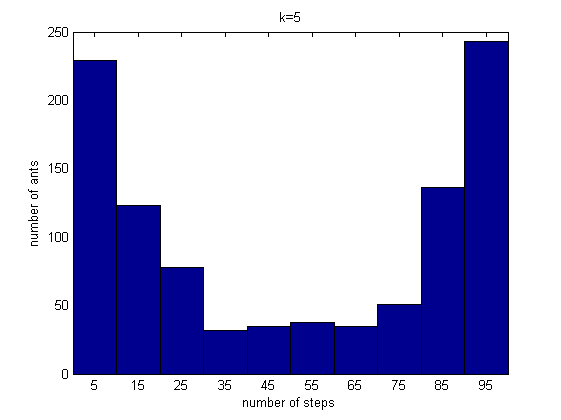}
\caption{Histogram for $k=5$, $\a=2$, $n=100$ time steps. The experiment is repeated $1000$ times and we represent the cumulative result.}
 \label{simu2}
 \end{figure}

\begin{figure} [h!]
\centering
    \includegraphics[width=0.75\textwidth]{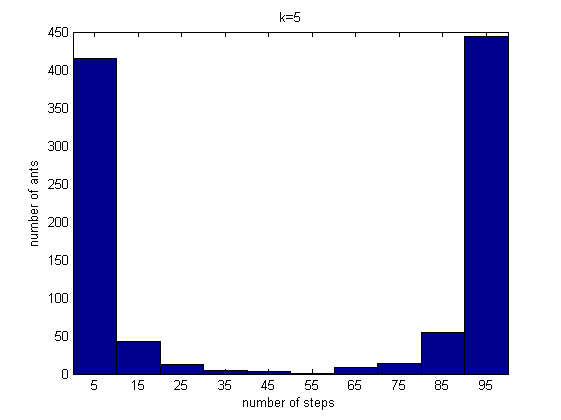}
\caption{Histogram for $k=5$, $\a=3$, $n=100$ time steps. The experiment is repeated $1000$ times and we represent the cumulative result.}
 \label{simu3}
 \end{figure}

\begin{figure} [h!]
\centering
    \includegraphics[width=0.75\textwidth]{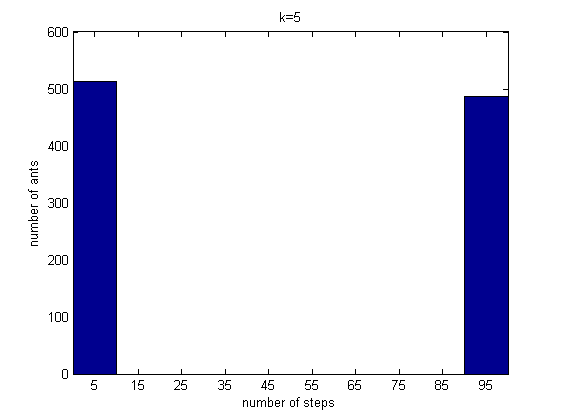}
\caption{Histogram for $k=5$, $\a=10$, $n=100$ time steps. The experiment is repeated $1000$ times and we represent the cumulative result.}
 \label{simu4}
 \end{figure}

\begin{figure} [h!]
\centering
    \includegraphics[width=0.75\textwidth]{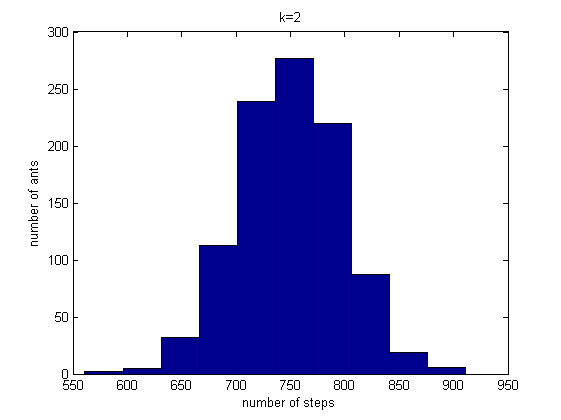}
\caption{Histogram for $k=2$, $\a=0.5$, $n=1000$ time steps. The experiment is repeated $1000$ times and we represent the cumulative result.}
 \label{simu5}
 \end{figure}

\begin{figure} [h!]
\centering
    \includegraphics[width=0.75\textwidth]{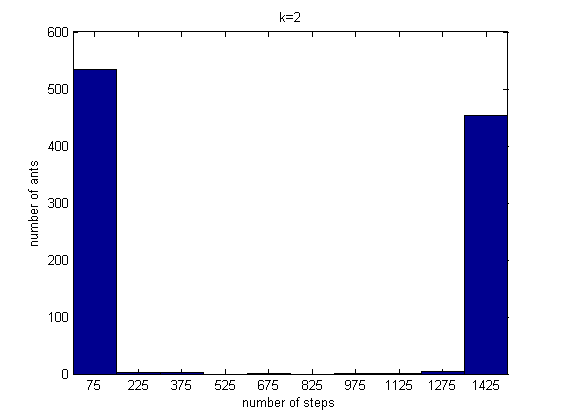}
\caption{Histogram for $k=2$, $\a=2$, $n=1500$ time steps. The experiment is repeated $1000$ times and we represent the cumulative result.}
 \label{simu6}
 \end{figure}

We perform numerical experiments for different sets of parameters ($k$, $\a$) to create histograms with
the behavior of one ant as a a function of the parameters.
In figures $\ref{simu1}-\ref{simu6}$ we consider one of the edges, say $W_1$ and a given number $n$ of time steps. We repeat the experiment a certain number of times and plot the number of this experiments for which the edge $W_1$ has been crossed by the ants a given number of times.
We can conclude from the graphics that
\begin{itemize}
\item [a)] If the ratio $k/\a$ is small both branches are selected equivalently; i.e. the ant chooses
one branch at the beginning and it chooses almost all times this branch.
\item [b)] If we increase the ratio $k/\a$ then the distribution becomes Gaussian.
\item [c)] For a fixed value of $k$, the larger is $\a$, the faster the histogram tends to a polarized state. These results agree with the rigorous mathematical results in \cite{Da} and \cite{V} showing that for $\a>1$ there is a path selection whereas for $\a<1$ there is not preferentiability.

%For a given value of $\a$, $k$ plays the role of time scaling and so the greater the value of $k$, the more polarized are the histograms (almost %all the ants go to the same branch). On the other hand, for a fixed value of $k$, the greater gets $\a$, the faster it tends to a polarized %state.

\end{itemize}

%---------------- INSERTAR FIGURA DE LA RED ---------------------------%
%--------------------------------------------------------------------%

%---------------- INSERTAR FIGURA DE LA RED ---------------------------%
%--------------------------------------------------------------------%

 %---------------- INSERTAR FIGURA DE LA RED ---------------------------%
%--------------------------------------------------------------------%

%---------------- INSERTAR FIGURA DE LA RED ---------------------------%
%--------------------------------------------------------------------%

%---------------- INSERTAR FIGURA DE LA RED ---------------------------%
%--------------------------------------------------------------------%

%---------------- INSERTAR FIGURA DE LA RED ---------------------------%
%--------------------------------------------------------------------%

%%%%%%%%%%%%%%%%%%%%%%%%%%%%%%%%%%%%%%%%%%%%%%%%%%%%%%%%%%%%%%%%%%%%%%%
\subsection{Simulations for a three node network}
%%%%%%%%%%%%%%%%%%%%%%%%%%%%%%%%%%%%%%%%%%%%%%%%%%%%%%%%%%%%%%%%%%%%%%%

In this section we consider a three node network as in figure \ref{network} right.
We distinguish two cases: non-constrained and directionality constrained.
In the constrained case, we will impose the following:
if the ant is at node $1$ or $2$, it can move to the other two nodes; if the ant
is at node $3$ and the previous node is node $1$, then it must move to node $2$; if
the previous node is node $2$, then it must move to node $1$.

For the case of a three node network with reinforcement (both with and without directionality constraint) we have four different states for the system: being at node $1$ (with associated probability $p_1$), being at node $2$ (with associated probability $p_2$), being at node $3$ coming from node $1$ (with associated probability $p_{3\uparrow 1}$) and being at node $3$ coming from node $2$ (with associated probability $p_{3\downarrow 2}$).
The transition probabilities are $p_{i,j}$ probability of moving from node $i$ to node $j$.
If we denote by $\om_i(t)$ the quantity of pheromone at link $W_i$ at time $t$ ($i=1,2,3)$, then the probabilities at step $t$ are given by:
\begin{equation}\label{prob:red3:eq1}
p_{1,2}(t)=\frac{(k+\om_1(t))^\a}{(k+\om_1(t))^\a+(k+\om_2(t))^\a},
\end{equation}
\begin{equation}\label{prob:red3:eq2}
p_{1,3}(t)=\frac{(k+\om_2(t))^\a}{(k+\om_1(t))^\a+(k+\om_2(t))^\a},
\end{equation}
\begin{equation}\label{prob:red3:eq3}
p_{2,1}(t)=\frac{(k+\om_1(t))^\a}{(k+\om_1(t))^\a+(k+\om_3(t))^\a},
\end{equation}
\begin{equation}\label{prob:red3:eq4}
p_{2,3}(t)=\frac{(k+\om_3(t))^\a}{(k+\om_1(t))^\a+(k+\om_3(t))^\a},
\end{equation}
\begin{equation}\label{prob:red3:eq5}
p_{3,1}(t)=\frac{(k+\om_2(t))^\a}{(k+\om_2(t))^\a+(k+\om_3(t))^\a},
\end{equation}
\begin{equation}\label{prob:red3:eq6}
p_{3,2}(t)=\frac{(k+\om_3(t))^\a}{(k+\om_2(t))^\a+(k+\om_3(t))^\a}.
\end{equation}

In the case with directionality constraint, since an ant in node $3$ is forced to go
to node $2$ if it is coming from node $1$ and to node $1$ if it is coming from node $2$, one must take
$p_{3,1}=1$ and $p_{3,2}=1$.

\begin{figure} [h!]
\centering
    \includegraphics[width=0.7 \textwidth]{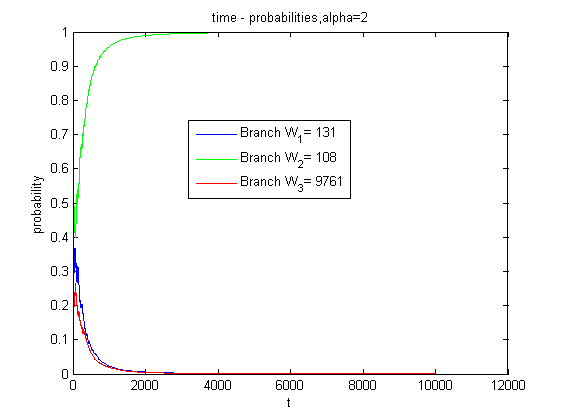}
    \caption{Temporal evolution for $n=10000$ time steps, $\a=2$ and $k=20$, network without directionality constraint. If $\a$ is sufficiently large, any branch can be selected.} %\vspace{-0.8cm}
    \label{simu14}
\end{figure}

\begin{figure} [h!]
\centering
    \includegraphics[width=0.7 \textwidth]{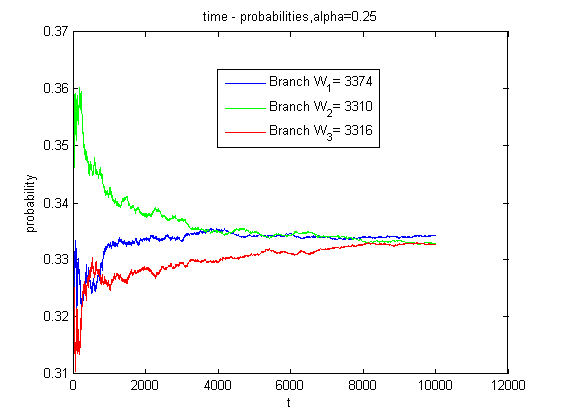}
    \caption{Temporal evolution for $n=10000$ time steps, $\a=0.25$ and $k=20$, network without directionality constraint. If $\a$ is small enough, no particular branch is selected.} %\vspace{-0.8cm}
    \label{simu15}
\end{figure}

\begin{figure} [h!]
\centering
    \includegraphics[width=0.69 \textwidth]{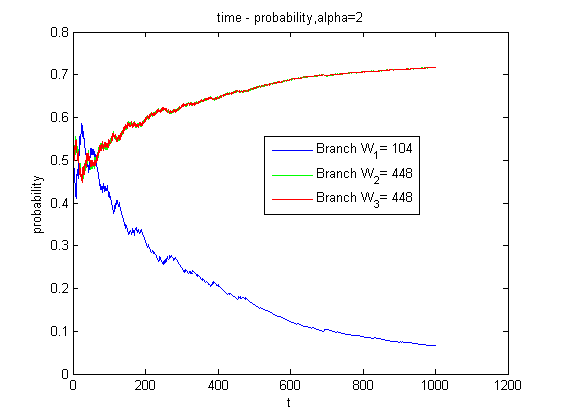}
    \caption{Temporal evolution for $n=1000$ time steps, $\a=2$ and $k=20$; network with directionality constraint.
For $\a$ sufficiently large one of the branches can be selected. In particular, with only one ant it may be selected the longest path.} %\vspace{-0.8cm}
    \label{simu16}
\end{figure}

\begin{figure} [h!]
\centering
    \includegraphics[width=0.69 \textwidth]{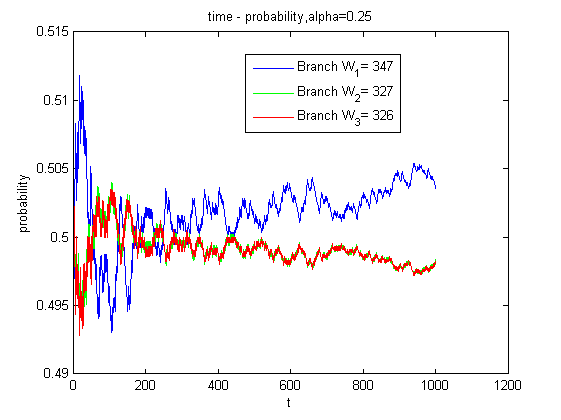}
    \caption{Temporal evolution for $n=1000$ time steps, $k=20$ and $\a=0.25$; network with directionality constraint.
If $\a$ is small enough, no particular branch is selected.} %\vspace{-0.8cm}
    \label{simu17}
\end{figure}

\begin{figure} [h!]
\centering
   \includegraphics[width=0.65 \textwidth]{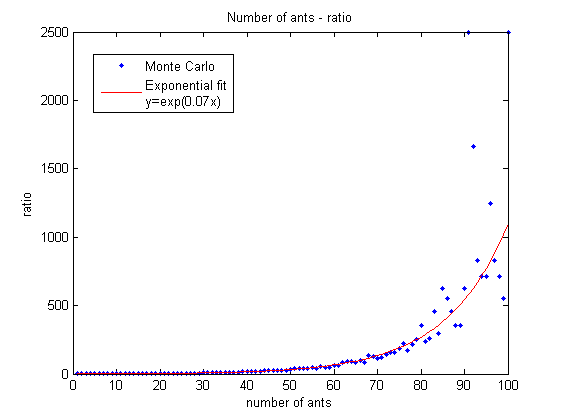}
    \caption{Ratio$=\om_1/\om_2$ for the number of times that the short path is chosen with respect to the long path as a function of the number of ants. The parameters are $k=20$, $\a=3$. Notice that the relative number of times that the shortest path has been selected grows exponentially with the number of ants: $y= \exp(0.07x)$.} %\vspace{-0.8cm}
    \label{simu26}
\end{figure}

\begin{figure} [h!]
\centering
    \includegraphics[width=0.65 \textwidth]{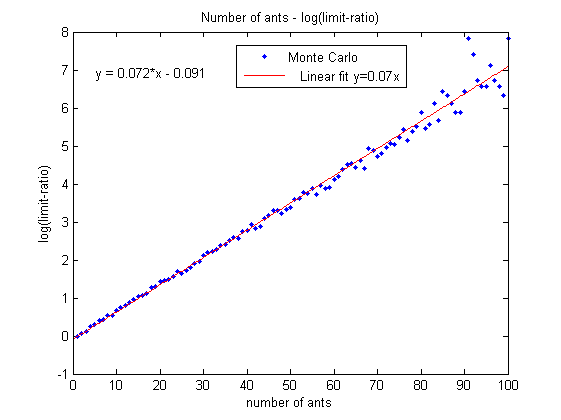}
    \caption{Logarithmic Ratio$=\om_1/\om_2$ for the number of times that the short path is chosen with respect to the long path as a function of the number of ants. The parameters are $k=20$, $\a=3$. Notice that the relative number of times that the shortest path has been selected grows exponentially with the number of ants: $y= \exp(0.07x)$.}
    \label{simu27}
\end{figure}

We employ different program simulations to show:
\begin{itemize}
\item [a)] Temporal evolution graphics to show the relative number of times one ant goes through each branch without the directionality constraint. We conclude that reinforcement is not enough to obtain selection of the shortest paths, since for $\a$ sufficiently large, one particular edge is reinforced but it is not necessarily the shortest one (see figure \ref{simu14}). If $\a$ is small enough, no particular branch is selected (see figure \ref{simu15}).

\item [b)] Temporal evolution graphics to show the number of times one ant goes through each branch with the directionality constraint.
We conclude that directionality constraint, with $\a$ sufficiently large, is sufficient to reinforce one particular path but this path is not necessarily the shortest one (see figure \ref{simu16}). If $\a$ is small enough, no particular branch is selected (see figure \ref{simu17}).

% Introducir las graficas del programa hormiga_v4
\item [c)] We consider more than one ant. For a certain number of numerical experiments $n$, we count for each one the relative number of times that the shortest path is chosen with respect to the number of times that the longest path is chosen. The result is a set of $n$ numbers at each time $t$ that we will call $r_i(t), i=1,\ldots,n,$ so that if $r_i(t)>1$, then for the i-th experiment at time $t$ the shortest path has been chosen most times and, on the contrary, if $r_i(t)<1$ then the longest path has been chosen most times. Then, at any given time $t$, we count the number of cases among the $n$ experiments for which the shortest path has been chosen (that is, those experiments for which $r_i(t)>1$) and divide it by the number of times that the longest path has been chosen (that is, those experiments for which $r_i(t)<1$). The result of these calculations is a measurement of the preferentiability of the shortest path with respect to the longest path. As we can see, the shortest path is chosen more often than the longest path for any number of ants.
    Notice that the convergence to a constant ratio is faster when the number of ants increases. In figure \ref{simu26} we represent the limiting ratio as a function of the number of ants and in figure \ref{simu27} its logarithm. As we can see, the ratio clearly follows an exponential law as a function of the number of ants, implying a strong reinforcement of shortest paths when the number of ants is relatively large.

\end{itemize}

\newpage
%---------------- INSERTAR FIGURA DE LA RED ---------------------------%
%---------------------------------------------------------------------%

%---------------- INSERTAR FIGURA DE LA RED ---------------------------%
%-----------------------------------------

%---------------- INSERTAR FIGURA DE LA RED ---------------------------%
%---------------------------------------------------------------------%

%---------------- INSERTAR FIGURA DE LA RED ---------------------------%
%---------------------------------------------------------------------%

%---------------- INSERTAR FIGURA DE LA RED ---------------------------%
%---------------------------------------------------------------------%

%---------------- INSERTAR FIGURA DE LA RED ---------------------------%
%---------------------------------------------------------------------%

As a result of the numerical simulations presented above, we conclude that reinforcement, persistence and a relatively large (in fact, more than one) number of ants are necessary for shortest selection in our three node network. The effect is stronger for large values of $\a$ (stronger nonlinearities) and increasing number of ants.
In fact, the number of times that the shortest path is selected relative to the number of times the longest path is selected grows exponentially with the number of ants (see figures \ref{simu26} and \ref{simu27}) implying that a large number of ants will find the shortest path quickly.

%%%%%%%%%%%%%%%%%%%%%%%%%%%%%%%%%%%%%%%%%%%%%%%%%%%%%%%%%%%%%%%%%%%%%%%
\section{Analytical results: long time dynamics}
%%%%%%%%%%%%%%%%%%%%%%%%%%%%%%%%%%%%%%%%%%%%%%%%%%%%%%%%%%%%%%%%%%%%%%%
In this section we discuss the possible dynamics at long times for the motion of ants in the networks represented in figure \ref{network}.

%%%%%%%%%%%%%%%%%%%%%%%%%%%%%%%%%%%%%%%%%%%%%%%%%%%%%%%%%%%%%%%%%%%%%%%
\subsection{Network with two nodes}
%%%%%%%%%%%%%%%%%%%%%%%%%%%%%%%%%%%%%%%%%%%%%%%%%%%%%%%%%%%%%%%%%%%%%%%

We consider a two node network with reinforcement as in figure \ref{network} left. Ants move one step at each time interval $\D t$ that can be taken, without loss of generality, as $\D t=1$. The probabilities are given by equations \eqref{prob:red2:eq1} and \eqref{prob:red2:eq2}.

We recall that $\D t$ is the time between two consecutive steps. We perform now a quasi-stationary approximation in the spirit of \cite{KSV} which consists in the following. Suppose that $t\gg 1$, then since $\omega_i$ is reinforced at each time step, $\omega_i$ is set of order $t$. Let us assume now that the ants perform $N$ steps with $N \D t\gg 1$ and $N \D t \ll t$. Since the characteristic time $\omega_i$ is of order $t$, we have that in the $N$ iterations the different amounts of pheromone $\omega_i$ can be assumed to be frozen. Therefore the evolution of the ants can be described with a markov chain with constant probabilities. Hence in times larger than $N \D t$ the occupancy times of the different nodes are proportional to the equilibrium probabilities for the markov chain. Since the size of the networks is of order one, the number of iterations needed for the system to approach equilibrium is of order one, and therefore the system can be assume to be at the equilibrium.

We can then compute the rate of change of the $\omega_i$ using these equilibrium distributions:
\[
\om_1(t+N \D t)-\om_1(t)=Np_{W_1},
\]
\[
\om_2(t+N \D t)-\om_2(t)=Np_{W_2},
\]
so that
\begin{equation}\label{ecu}
\left \{ \begin{array}{c}
\begin{split}
\frac{\om_1(t+N \D t)-\om_1(t)}{N}=p_{W_1}, \\
\frac{\om_2(t+N \D t)-\om_2(t)}{N}=p_{W_2}.
\end{split}
\end{array} \right.
\end{equation}

Asymptotically, using our choice of $\D t=1$, we replace the left hand side of \eqref{ecu} by time derivatives and then
\begin{equation}\label{siste}
 \left \{ \begin{array}{c}
 \begin{split}
  \frac{d\om_1}{dt}&= p_{W_1},\\
  \frac{d\om_2}{dt}&= p_{W_2},
  \end{split}
\end{array} \right.
\end{equation}
with $p_{W_1}+p_{W_2}=1$, since we are working with probabilities, and $\om_1+\om_2=t$.

If we do the change $\om_i=k \Om_i, i=1,2, t=k \t$ then system (\ref{siste}) becomes
\begin{equation}\label{siste2}
 \left \{ \begin{array}{c}
 \begin{split}
  \frac{d\Om_1}{d\t}&=\frac{(1+\Om_1)^\a}{(1+\Om_1)^\a+(1+\Om_2)^\a},\\
  \frac{d\Om_2}{d\t}&=\frac{(1+\Om_2)^\a}{(1+\Om_1)^\a+(1+\Om_2)^\a},
  \end{split}
\end{array} \right.
\end{equation}
and $\Om_1+\Om_2=\t$.

Now, in order to study the equilibrium points for system \eqref{siste2}, we perform the change $\p_i=\frac{\Om_i}{\t},i=1,2$.
Since $\p_1+\p_2=1$, we only need to take into account branch $\p_1$. Hence

\begin{equation}\label{ecu:equil}
\frac{d\Om_1}{d\t}=\frac{d}{d\t}(\t \p_1)=\p_1+\t\frac{d\p_1}{d\t}=  \frac{(1+\t\p_1)^\a}{(1+\t\p_1)^\a+(1+\t\p_2)^\a},
\end{equation}
and so
\begin{equation}\label{ecu:equil2}
\t\frac{d\p_1}{d\t}=\frac{1}{1+\Big(1-\frac{2\p_1-1}{\p_1+\frac{1}{\t}}\Big)^\a}-\p_1.
\end{equation}

If $1\ll\t,$ and as long as $\p_1$ is of order one, linearizing in \eqref{ecu:equil2} and performing the change $\eta=\log(\t)$ we have
\begin{equation}\label{EDO2}
\boxed {\frac{d\p_1}{d\eta}=\frac{1}{1+\Big(\frac{1}{\p_1}-1 \Big)^\a}-\p_1 .}
\end{equation}

The equilibria of \eqref{EDO2} are:
\[
\p_1=1,\qquad \p_1=0,\qquad \p_1=\frac 12.
\]

Hence, the equilibrium points are
\[
(1,0),\qquad (0,1),\qquad (\frac 12,\frac 12).
\]

We study in detail the behavior at each equilibrium point.

\begin{description}
\item [\underline{Case $\p_1=\frac 12$}.]

We consider the approximation
\[
\p_1=\frac 12 + \tilde{\p_1}.
\]
Introducing this value into equation \eqref{EDO2} we have
\begin{equation*}
\frac{d\tilde{\p_1}}{d\eta}\approx (\a-1)\tilde{\p_1},
\end{equation*}
where we have used Taylor's expansions.

Then, we have two different cases:
\begin{itemize}
\item [a)] If $\a<1$,
\[
\frac{d\tilde{\p_1}}{d\eta}=\overbrace{(\a-1)}^{<0}\tilde{\p_1}\Rightarrow \tilde{\p_1}\approx C e^{(-|\a-1|\eta)}.
\]
$\p_1=\frac 12$ is STABLE.
\item [b)] If $\a>1$,
\[
\frac{d\tilde{\p_1}}{d\eta}=\overbrace{(\a-1)}^{>0}\tilde{\p_1}\Rightarrow \tilde{\p_1}\approx C e^{(|\a-1|\eta)}.
\]
$\p_1=\frac 12$ is UNSTABLE.
\end{itemize}

\item [\underline{Case $\p_1=1$}.]

We consider the approximation
\[
\p_1=1- \tilde{\p_1}.
\]
Introducing this value into equation \eqref{EDO2} we have
\begin{equation*}
-\frac{d\tilde{\p_1}}{d\eta}\approx \tilde{\p_1}-\tilde{\p_1}^\a,
\end{equation*}
where we have used Taylor's expansions.

We have the following cases:
\begin{itemize}
\item [a)] If $\a>1$, since $\tilde{\p_1}>\tilde{\p_1}^\a$,then
\[
\frac{d\tilde{\p_1}}{d\eta}=-\tilde{\p_1}\Rightarrow \tilde{\p_1}\approx C e^{-\eta}.
\]
$\p_1=1$ is STABLE.
\item [b)] If $\a<1$, since $\tilde{\p_1}<\tilde{\p_1}^\a$, then
\[
\frac{d\tilde{\p_1}}{d\eta}=\tilde{\p_1}^\a\Rightarrow \tilde{\p_1}\approx C \eta^{\frac{1}{1-\a}}.
\]
$\p_1=1$ is UNSTABLE.
\end{itemize}

\item [\underline{Case $\p_1=0$}.]

We consider the approximation
\[
\p_1= \tilde{\p_1}.
\]
Introducing this value into equation \eqref{EDO2} we have
\begin{equation*}
\frac{d\tilde{\p_1}}{d\eta}\approx \tilde{\p_1}^\a-\tilde{\p_1}.
\end{equation*}
where we have used Taylor's expansions.

We have the following cases:
\begin{itemize}
\item [a)] If $\a>1$, since $\tilde{\p_1}>\tilde{\p_1}^\a$, then
\[
\frac{d\tilde{\p_1}}{d\eta}=-\tilde{\p_1}\Rightarrow \tilde{\p_1}\approx C e^{-\eta}.
\]
$\p_1=0$ is STABLE.
\item [b)] If $\a<1$, since $\tilde{\p_1}<\tilde{\p_1}^\a$, then
\[
\frac{d\tilde{\p_1}}{d\eta}=\tilde{\p_1}^{\a}\Rightarrow \tilde{\p_1}\approx C \eta^{\frac{1}{1-\a}}.
\]
$\p_1=0$ is UNSTABLE.
\end{itemize}

\end{description}

Therefore, $\a$ appears as a critical parameter for reinforcement of edges. If $\a<1$ then non-reinforcement will take place since the state $(\p_1,\p_2)=(\frac 12, \frac 12)$ is stable. On the other hand, if $\a>1$ then one edge or the other will be reinforced since both $(\p_1,\p_2)=(1,0)$ and $(\p_1,\p_2)=(0,1)$ become stable. The result, of course, supports the numerical observations in the previous section.

%%%%%%%%%%%%%%%%%%%%%%%%%%%%%%%%%%%%%%%%%%%%%%%%%%%%%%%%%%%%%%%%%%%%%%%%%%%%%%%%%%%
\subsection{Network with three nodes}
%%%%%%%%%%%%%%%%%%%%%%%%%%%%%%%%%%%%%%%%%%%%%%%%%%%%%%%%%%%%%%%%%%%%%%%%%%%%%%%%%%%

We consider a three node network as in figure \ref{network} right with reinforcement.
With the same directionality constraint as in the previous section, the probabilities for each state are given by equations \eqref{prob:red3:eq1}, \eqref{prob:red3:eq2}, \eqref{prob:red3:eq3} and \eqref{prob:red3:eq4}.

For $k\gg 1,t \gg k,t\gg N \D t$, at a time scale $[t,t+N \D t]$, that is under the hypothesis for the quasi-stationary approximation done in the case of the two node network, we have that:

\[
\om_1(t+N \D t)-\om_1(t)=N(p_{2,1}p_2+p_{1,2}p_1),
\]
\[
\om_2(t+N \D t)-\om_2(t)=N(p_{1,3}p_1+p_{3\downarrow  2}),
\]
\[
\om_3(t+N \D t)-\om_3(t)=N(p_{2,3}p_2+p_{3\uparrow 1}),
\]
where the $p_1, p_2, p_{3\downarrow 2}$ and $p_{3\uparrow 1}$ are at equilibrium.

Asymptotically, taking $\D t=1$ and approximating $\frac {1}{ N}(\om_i(t+N \D t)-\om_i(t))\approx \frac{d\om_i}{dt}, i=1,2,3$, we get
\begin{equation}\label{sistema}
 \left \{ \begin{array}{c}
 \begin{split}
  \frac{d\om_1}{dt}&= p_{2,1}p_2+p_{1,2}p_1,\\
  \frac{d\om_2}{dt}&= p_{1,3}p_1+p_{3\downarrow 2}, \\
  \frac{d\om_3}{dt}&= p_{2,3}p_2+p_{3\uparrow 1},
  \end{split}
\end{array} \right.
\end{equation}
with $p_{1,3}p_1+p_{3\downarrow2}+p_{2,1}p_2+p_{1,2}p_1+p_{2,3}p_2+p_{3\uparrow 1}=1$, since we are working with probabilities, and $\om_1+\om_2+\om_3=t$.

The equations for the transition probabilities of the four different states are then
\begin{equation}\label{sistema:prob}
 \left \{ \begin{array}{c}
 \begin{split}
  p_{1}&= p_{2}\frac{(k+\om_1)^\a}{(k+\om_1)^\a+(k+\om_3)^\a}+p_{3\downarrow 2},\\
  p_{2}&= p_{1}\frac{(k+\om_1)^\a}{(k+\om_1)^\a+(k+\om_2)^\a}+p_{3\uparrow 1}, \\
  p_{3\uparrow 1}&= p_{1}\frac{(k+\om_2)^\a}{(k+\om_1)^\a+(k+\om_2)^\a},\\
  p_{3\downarrow 2}&= p_{2}\frac{(k+\om_3)^\a}{(k+\om_1)^\a+(k+\om_3)^\a}.
  \end{split}
\end{array} \right.
\end{equation}
If we do the change $\om_i=k \Om_i, i=1,2,3, t=k \t$ then system (\ref{sistema}) becomes
\begin{equation}\label{sistema2}
 \left \{ \begin{array}{c}
 \begin{split}
  \frac{d\Om_1}{d\t}&= \frac{(k\Om_1+k)^\a}{(k\Om_1+k)^\a+(k\Om_3+k)^\a}p_2\\
  &\qquad +\frac{(k\Om_1+k)^\a}{(k\Om_1+k)^\a+(k\Om_2+k)^\a}p_1,\\
  \frac{d\Om_2}{d\t}&= \frac{(k\Om_2+k)^\a}{(k\Om_1+k)^\a+(k\Om_2+k)^\a}p_1+p_{3\downarrow 2}, \\
  \frac{d\Om_3}{d\t}&= \frac{(k\Om_3+k)^\a}{(k\Om_1+k)^\a+(k\Om_3+k)^\a}p_2+p_{3\uparrow 1},
  \end{split}
\end{array} \right.
\end{equation}
and $\Om_1+\Om_2+\Om_3=\t$.

Similarly, for $N\gg 1$, system \eqref{sistema:prob} becomes

\begin{empheq}[left=\empheqlbrace]{align} %Para poder poner etiquetas a cada ecuaci\'{o}n
p_{1}  &= p_{2}\frac{\Om_1^\a}{\Om_1^\a+\Om_3^\a}+p_{3\downarrow 2},  \label{eqprob:1}\\
p_{2}  &= p_{1}\frac{\Om_1^\a}{\Om_1^\a+\Om_2^\a}+p_{3\uparrow 1},  \label{eqprob:2}\\
p_{3\uparrow 1}  &= p_{1}\frac{\Om_2^\a}{\Om_1^\a+\Om_2^\a}, \label{eqprob:3}\\
p_{3\downarrow 2}  &= p_{2}\frac{\Om_3^\a}{\Om_1^\a+\Om_3^\a}. \label{eqprob:4}
\end{empheq}
Since $\t\gg 1$, then by writing $\Om_i=\p_i\t, i=1,2,3$, asymptotically the system (\ref{sistema2}) becomes
\begin{empheq}[left=\empheqlbrace]{align} %Para poder poner etiquetas a cada ecuaci\'{o}n
\p_1  &= \frac{\p_1^\a}{\p_1^\a+\p_3^\a}p_2+\frac{\p_1^\a}{\p_1^\a+\p_2^\a}p_1,  \label{ec:1}\\
\p_2  &= \frac{\p_2^\a}{\p_1^\a+\p_2^\a}p_1+p_{3\downarrow 2}, \label{ec:2}\\
\p_3  &= \frac{\p_3^\a}{\p_1^\a+\p_3^\a}p_2+p_{3\uparrow 1}, \label{ec:3}
\end{empheq}
and system \eqref{eqprob:1}-\eqref{eqprob:4} holds with $\Om_i$ replaced by $\p_i$:
\begin{empheq}[left=\empheqlbrace]{align} %Para poder poner etiquetas a cada ecuaci\'{o}n
p_{1}  &= p_{2}\frac{\p_1^\a}{\p_1^\a+\p_3^\a}+p_{3\downarrow 2},  \label{ecprob:1}\\
p_{2}  &= p_{1}\frac{\p_1^\a}{\p_1^\a+\p_2^\a}+p_{3\uparrow 1},  \label{ecprob:2}\\
p_{3\uparrow 1}  &= p_{1}\frac{\p_2^\a}{\p_1^\a+\p_2^\a}, \label{ecprob:3}\\
p_{3\downarrow 2}  &= p_{2}\frac{\p_3^\a}{\p_1^\a+\p_3^\a}. \label{ecprob:4}
\end{empheq}

Substituting \eqref{ecprob:3} into \eqref{ecprob:2} we get
\begin{equation}\label{ec:igualprob}
p_{1}\frac{\p_1^\a}{\p_1^\a+\p_2^\a}+p_{1}\frac{\p_2^\a}{\p_1^\a+\p_2^\a}-p_2=0 \Rightarrow \boxed {p_2=p_1,}
\end{equation}
and introducing \eqref{ecprob:3}, \eqref{ecprob:4}, \eqref{ec:igualprob} into \eqref{ec:2} and \eqref{ec:3} we have

\begin{equation}\label{ec:igualome}
\left \{ \begin{array}{c}
 \begin{split}
  \p_2&=\frac{\p_2^\a}{\p_1^\a+\p_2^\a}p_1+ \frac{\p_3^\a}{\p_1^\a+\p_3^\a}p_{1}, \\
  \p_3&=\frac{\p_3^\a}{\p_1^\a+\p_3^\a}p_1+ \frac{\p_2^\a}{\p_1^\a+\p_2^\a}p_{1},
  \end{split}
\end{array} \right. \Rightarrow \boxed {\p_2=\p_3.}
\end{equation}

By substituting \eqref{ec:igualome} and \eqref{ec:igualprob} into \eqref{ecprob:3} and \eqref{ecprob:4} we have
\begin{equation}\label{ec:igualprob2}
 \boxed {p_{3\downarrow 2}=p_{3\uparrow 1}.}
\end{equation}

Finally, introducing \eqref{ec:igualprob} and \eqref{ec:igualome} into \eqref{ecprob:2} we have
\begin{equation}\label{ec:probOmega}
 p_1\Bigg( 1-\frac{(1-2\p_2)^\a}{(1-2\p_2)^\a+\p_2^\a}\Bigg)=p_{3\uparrow 1}.
\end{equation}

Since $p_1+p_2+p_{3\uparrow 1}+p_{3\downarrow 2}=1$ and $p_1=p_2,p_{3\uparrow 1}=p_{3\downarrow 2}$ we have

\begin{equation}\label{ec:probIg}
 \frac 12-p_1=p_{3\uparrow 1}.
\end{equation}

Plugging \eqref{ec:probIg} into \eqref{ec:probOmega} we get
\begin{equation}\label{ec:probOmega2}
 p_1\Bigg( 1-\frac{(1-2\p_2)^\a}{(1-2\p_2)^\a+\p_2^\a}\Bigg)=\frac 12-p_1\Rightarrow p_1=
 \frac 12\Bigg(\frac{1}{1+\frac{\p_2^\a}{\p_2^\a+\p_1^\a}} \Bigg).
\end{equation}

Now, we study the equilibrium points for system \eqref{sistema2} as well as their stability. Since $\p_1+\p_2+\p_3=1$ and $\p_2=\p_3$ by \eqref{ec:igualome}, we only take into account the equation for $\p_1$. Hence

\begin{equation}\label{ec:equil}
\frac{d\Om_1}{d\t}=\p_1+\t\frac{d\p_1}{d\t}=2 p_1 \Bigg( \frac{(1+\t\p_1)^\a}{(1+\t\p_1)^\a+(1+\t\p_2)^\a}\Bigg).
\end{equation}

Introducing the value of $p_1$ obtained in \eqref{ec:probOmega2} into \eqref{ec:equil} and approximating for $\t \gg 1$ we get

\begin{equation}\label{ec:equil2}
\t\frac{d\p_1}{d\t} \approx \frac{1}{1+2\Big(\frac{\p_2}{\p_1}\Big)^\a}-\p_1 =\frac{1}{1+2\Big(\frac{\frac{1-\p_1}{2}}{\p_1}\Big)^\a}-\p_1.
\end{equation}
where we have used that $\p_2=\frac{1-\p_1}{2}$.

Therefore, \eqref{ec:equil2} becomes

\begin{equation}\label{ec:equil3}
\t\frac{d\p_1}{d\t}= \frac{1}{1+2^{(1-\a)}(-1+\p_1^{-1})^\a}-\p_1,
\end{equation}
which provides an equation for $\p_1$ provided that $1\ll\t,\p_1\gg \t^{-1}$.

If we do the change $\eta=\log(\t)$ in \eqref{ec:equil3} we have
\begin{equation}\label{EDO}
\boxed {\frac{d\p_1}{d\eta}=\frac{1}{1+2^{(1-\a)}\Big(\frac{1}{\p_1}-1 \Big)^\a}-\p_1. }
\end{equation}

To find the equilibrium points we calculate
\[
\frac{d\p_1}{d\eta}=0\Leftrightarrow (2\p_1)^{(\a-1)}=(2\p_1)^{(\a-1)}\p_1+(1-\p_1)^\a,
\]
and by straightforward calculations we get

\[
\p_1=1,\qquad \p_1=0,\qquad \p_1=\frac 13.
\]

Hence, the equilibrium points are
\[
(1,0,0),\qquad (0,\frac 12,\frac 12),\qquad (\frac 13,\frac 13,\frac 13).
\]

We study in detail the behavior at each equilibrium point.

\begin{description}
\item [\underline{Case $\p_1=\frac 13$}.]

We consider the approximation
\[
\p_1=\frac 13 + \tilde{\p_1}.
\]
Introducing this value into equation \eqref{EDO} we have
\begin{equation*}
\frac{d\tilde{\p_1}}{d\eta}=\frac{1}{1+2\Bigg(\frac{\frac 23-\tilde{\p_1}}{\frac23+2\tilde{\p_1}}\Bigg)^\a}-\frac13-\tilde{\p_1}=(\a-1)\tilde{\p_1}.
\end{equation*}
where we have used Taylor's expansions.

Then, we have two different cases:
\begin{itemize}
\item [a)] If $\a<1$,
\[
\frac{d\tilde{\p_1}}{d\eta}=\overbrace{(\a-1)}^{<0}\tilde{\p_1}\Rightarrow \tilde{\p_1}\approx C e^{(-|\a-1|\eta)}.
\]
$\p_1=\frac 13$ is STABLE.
\item [b)] If $\a>1$,
\[
\frac{d\tilde{\p_1}}{d\eta}=\overbrace{(\a-1)}^{>0}\tilde{\p_1}\Rightarrow \tilde{\p_1}\approx C e^{(|\a-1|\eta)}.
\]
$\p_1=\frac 13$ is UNSTABLE.
\end{itemize}

\item [\underline{Case $\p_1=1$}.]

We consider the approximation
\[
\p_1=1- \tilde{\p_1}.
\]
Introducing this value into equation \eqref{EDO} we have
\begin{equation*}
-\frac{d\tilde{\p_1}}{d\eta}=\frac{1}{1+2\Bigg(\frac 12\frac {\tilde{\p_1}}{1-\tilde{\p_1}}\Bigg)^\a}-1+\tilde{\p_1}=-2^{1-\a}\tilde{\p_1}^\a+\tilde{\p_1},
\end{equation*}
where we have used Taylor's expansions.

We have the following cases:
\begin{itemize}
\item [a)] If $\a>1$, since $\tilde{\p_1}\gg\tilde{\p_1}^\a$, then
\[
\frac{d\tilde{\p_1}}{d\eta}=-\tilde{\p_1}\Rightarrow \tilde{\p_1}\approx C e^{-\eta}.
\]
$\p_1=1$ is STABLE.
\item [b)] If $\a<1$, since $\tilde{\p_1}\ll\tilde{\p_1}^\a$, then
\[
\frac{d\tilde{\p_1}}{d\eta}=2^{(1-\a)}\tilde{\p_1}^\a\Rightarrow \tilde{\p_1}\approx C \eta^{\frac{1}{1-\a}}.
\]
$\p_1=1$ is UNSTABLE.
\end{itemize}

\item [\underline{Case $\p_1=0$}.]

We consider $\p_1$ small.
Introducing this value into equation \eqref{EDO} we have
\begin{equation*}
\frac{d\p_1}{d\eta}=\frac{1}{1+2\Bigg(\frac {1-\p_1}{2\p_1}\Bigg)^\a}-\p_1=\approx 1-2^{(1-\a)}(1-\p_1^{-\a})-\p_1,
\end{equation*}
where we have used Taylor's expansions.

We have the following cases:
\begin{itemize}
\item [a)] If $\a>1$, since $\p_1\gg \p_1^\a$, then
\[
\frac{d\p_1}{d\eta}=-\p_1\Rightarrow \p_1\approx C e^{-\eta}.
\]
$\p_1=0$ is STABLE.
\item [b)] If $\a<1$, since $\p_1\ll\p_1^\a$, then
\[
\frac{d\p_1}{d\eta}=1-2^{(1-\a)}\p_1^{-\a}\Rightarrow \p_1\approx C \eta^{\frac{\a}{1+\a}}.
\]
$\p_1=0$ is UNSTABLE.
\end{itemize}

\end{description}

As a conclusion of the analysis of the three node network, if $\a<1$ then the three edges are run with identical probability
since the only stable equilibrium is $(\p_1,\p_2,\p_3)=(\frac 13,\frac 13,\frac 13)$, while for $\a>1$ the states $(\p_1,\p_2,\p_3)=(1,0,0)$ and $(\p_1,\p_2,\p_3)=(0, \frac 12, \frac 12)$, corresponding to the shortest and longest paths respectively, are the stable ones. This result implies that for $\a>1$ one particular path, the short or the long one, will be reinforced and our random walker will end up walking on it with a probability that tends to one as time goes to infinity. Nevertheless, the analysis does not provide a reason for the shortest one to be selected preferably with respect to the longest one. In the next section, we will see that such a selection takes place in the first stages of the evolution when reinforcement is still very weak and provided that more than one ant are running through the network. In order to perform this analysis, we will linearize the probabilities given by \eqref{prob:red3:eq1}-\eqref{prob:red3:eq4} using as a small parameter $\a/k$ and solve the resulting evolution problem. By assuming $\a/k\ll 1$ we are considering the case where reinforcement remains very weak up to times when $\om_i=\mathcal{O}(k)$. We will show that the difference in the amount of pheromone between the shortest path and any of the links in the longest path has a probability distribution that evolves according to a convection equation. The convection velocity, when there is more than one ant, is always in the direction of increasing the value of the difference in the amount of pheromone and hence the shortest path will be increasingly reinforced. This breaking of symmetry occurs faster with increasing number of ants due to the fact that the convection velocity grows very quickly with the number of ants.

%%%%%%%%%%%%%%%%%%%%%%%%%%%%%%%%%%%%%%%%%%%%%%%%%%%%%%%%%%%%%%%%%%%%%%%%%%%
\section{Analytical results: early time dynamics with weak reinforcement}
%%%%%%%%%%%%%%%%%%%%%%%%%%%%%%%%%%%%%%%%%%%%%%%%%%%%%%%%%%%%%%%%%%%%%%%%%%%%

%%%%%%%%%%%%%%%%%%%%%%%%%%%%%%%%%%%%%%%%%%%%%%%%%%%%%%%%%%%%%%%%%%%
\subsection{Reinforced and non-reinforced network with one ant}
%%%%%%%%%%%%%%%%%%%%%%%%%%%%%%%%%%%%%%%%%%%%%%%%%%%%%%%%%%%%%%%%%%%

Considering the three node network in figure \ref{network} right for one ant, we have two possible states:
\begin{itemize}
\item The ant is at food source (node 2) or nest (node 1), case $\mathcal{A}$.
\item The ant is at node $3$, case $\mathcal{B}$.
\end{itemize}

To simplify the analysis we restrict the problem to times $t$ so that $\om_i=\mathcal{O}(t) \ll \frac{k}{\a}$. This corresponds to the case where reinforcement is still very weak due to the fact that $\om_i \ll k$ in formulas \eqref{prob:red3:eq1}-\eqref{prob:red3:eq4}. We can then approximate the probability $p_{1,2}$ in formula \eqref{prob:red3:eq1} by

\[
\begin{split}
p_{1,2}&=\frac{(k+\om_1)^\a}{(k+\om_1)^\a+(k+\om_2)^\a}\approx \frac{1+\a\frac{\om_1}{k}}{2+\a\left( \frac{\om_1}{k}+\frac{\om_2}{k}\right)}\\
&\approx \frac 12 \left(1+\a\frac{\om_1-\om_2}{2k} \right)=\frac{1+\e\D}{2},
\end{split}
\]
where $\e=\frac{\a}{2k}$ and $\D=\om_1-\om_2$.

Analogously,
\[
p_{1,3}\approx \frac{1-\e\D}{2},\qquad p_{2,1}\approx \frac{1+\e\D}{2},\qquad p_{2,3}\approx \frac{1-\e\D}{2}.
\]
Notice that we have approximated $\om_2=\om_3$ since both edges $W_2,W_3$ are run the same number of times due to the directionality constraint imposed.
Considering the cases $\mathcal{A}$ and $\mathcal{B}$, we can describe any possible evolution as a sequence of the following states:
\begin{enumerate}
  \item From state $\mathcal{A}$ to state $\mathcal{A}$: $p=\frac{1+\e\D}{2}, \D\rightarrow \D+1$.
  \item From state $\mathcal{A}$ to state $\mathcal{A}$ passing through state $\mathcal{B}$: $p=\frac{1-\e\D}{2}, \D\rightarrow \D-1$.
\end{enumerate}

Then the master equation for the probability is:

\begin{equation}\label{prob1ant}
p(\D,N+1)= \frac 12 (1-\e\D) p(\D+1,N)+\frac 12(1+\e\D) p(\D-1,N).
\end{equation}

%%%%%%%%%%%%%%%%%%%%%%%%%%%%%%%%%%%%%%%%%%%%%%%%%%%%%%%%%%%%%%%%%%%
\emph{Equations at $\mathcal{O}(\varepsilon^0)$}
%%%%%%%%%%%%%%%%%%%%%%%%%%%%%%%%%%%%%%%%%%%%%%%%%%%%%%%%%%%%%%%%%%%

From \eqref{prob1ant} we have that

\begin{equation}\label{prob1ant2}
p(\D,N+1)= \frac 12  p(\D+1,N)+\frac 12 p(\D-1,N).
\end{equation}

If we subtract $p(\D,N)$ at both sides in \eqref{prob1ant2}, we get

\[
\frac{\partial p}{\partial t}\approx p(\D,N+1)-p(\D,N)\approx \frac 12 \frac{\partial^2 p}{\partial \D^2}.
\]

This is a diffusion equation without transport terms (i.e. terms involving $\frac{\partial p}{\partial x}$) and hence the solution is such that if $p(\D,0)$ is centered at $\D=\D_0$ then $p(\D,t)$ will also be centered at $\D=\D_0$. Therefore, the maximum probability will always be at $\D=\D_0$ and hence no path will be reinforced.

%%%%%%%%%%%%%%%%%%%%%%%%%%%%%%%%%%%%%%%%%%%%%%%%%%%%%%%%%%%%%%%%%%%
\emph{Equations at $\mathcal{O}(\varepsilon^1)$}
%%%%%%%%%%%%%%%%%%%%%%%%%%%%%%%%%%%%%%%%%%%%%%%%%%%%%%%%%%%%%%%%%%%

If we subtract $p(\D,N)$ at both sides in \ref{prob1ant}, we get

\[
\frac{\partial p}{\partial t}\approx \frac 12 \frac{\partial^2 p}{\partial \D^2} -\e\D \frac{\partial p}{\partial \D},
\]
where we have done the following approximations
\[
\frac{\partial p}{\partial t}\approx p(\D,N+1)-p(\D,N),
\]
\[
\frac{\partial p}{\partial \D}\approx \frac{p(\D+1,N)-p(\D-1,N)}{2},
\]
\[
\frac{\partial^2 p}{\partial \D^2}\approx p(\D+1,N)+p(\D-1,N)-2p(\D,N).
\]

If $p(\D,0)$ is centered at $\D=\D_0$ then the presence of a convective term with velocity $\e\D$ will produce a shift of the movement of $p(\D,t)$
to increasing (if $\D_0>0$) values of $\D$ or to decreasing (if $\D_0<0$) values of $\D$. Hence, one of the paths, the short or the long one, will be
reinforced depending on the initial condition. This agrees with our previous numerical simulations concerning the fact that only one ant is able to reinforce one of the paths but more than one ant is necessary to actually reinforce the shortest one preferably.

%%%%%%%%%%%%%%%%%%%%%%%%%%%%%%%%%%%%%%%%%%%%%%%%%%%%%%%%%%%%%%%%%%%
\subsection{Reinforced and non-reinforced network with two ants}
%%%%%%%%%%%%%%%%%%%%%%%%%%%%%%%%%%%%%%%%%%%%%%%%%%%%%%%%%%%%%%%%%%%

In this section, we consider at the same time both the reinforced and non-reinforced cases for two ants in a three node network with directionality constraints.

We show that it is enough to consider both directionality constraint and reinforcement to reproduce ant's behavior concerning choice of the shortest path.

Considering the three node network in figure \ref{network} right for two ants, we can classify any state into these four different states:
\begin{enumerate}
\item Both ants at nest (node 1) or food source (node 2), case $\mathcal{A^+}$;
\item One ant at nest and the other at food source, case $\mathcal{A^-}$;
\item Both ants at node $3$, case $\mathcal{B}$;
\item One ant at nest/food source and the other at node $3$, case $\mathcal{C}$.
\end{enumerate}

%---------------- INSERTAR FIGURA DE LA RED ---------------------------%
\begin{figure}[h!]
\centering
    \includegraphics[width=0.6\textwidth]{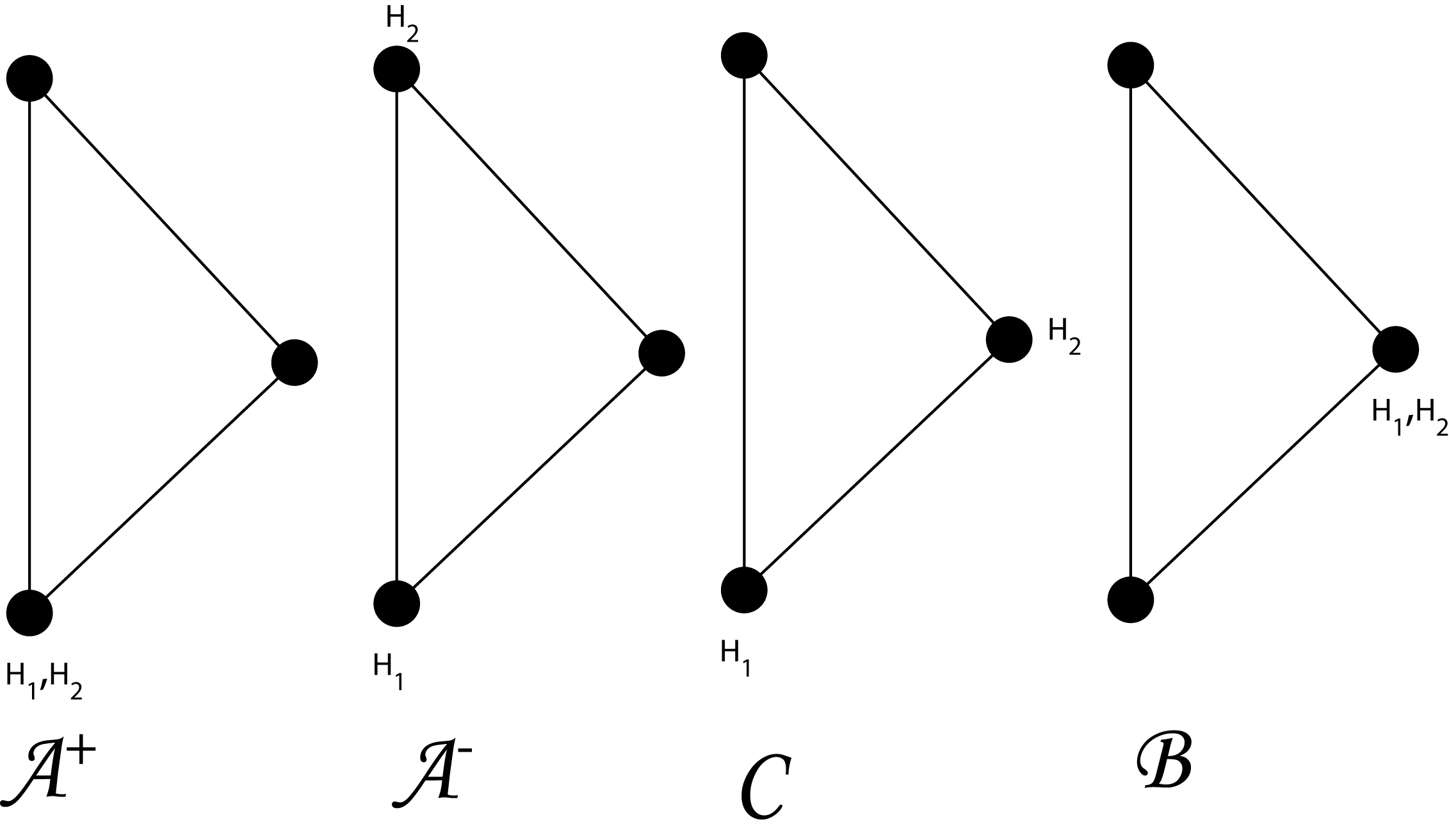}
    \caption{Different ant's states. State $\mathcal{A^+}$ also represents the situation where both ants are at the food source. The state $\mathcal{C}$ also represents the situation where one ant is at the food source and the other at node three.}
    \label{redes}
\end{figure}
%---------------------------------------------------------------------%

By employing these $4$ states, we can describe any possible evolution as a sequence of such states. The probability of being at one of the states at a certain time will depend on the probabilities of having previously been at other states. In order to arrive to a simple way to compute such probabilities, we find a representation of the evolution as a markovian process where one can write the probability to reach a certain state at time $N+1$ merely as a function of the probabilities to be at each state at time $N$. The directionality imposed in the problem reduces drastically the number of possible transitions from one state to the other so that any evolution of the system can be viewed as a sequence of "syllables"
\[
\mathcal{A^+}\mathcal{A^+},\qquad \mathcal{A^-}\mathcal{A^-},\qquad \mathcal{A^+}\mathcal{B}\mathcal{A^+},\qquad \mathcal{A^-}\mathcal{B}\mathcal{A^-},
\]
\[
\mathcal{A^+}\mathcal{C}^{(2j)}\mathcal{A^+},\mathcal{A^+}\mathcal{C}^{(2j+1)}\mathcal{A^-},
\mathcal{A^-}\mathcal{C}^{(2j+1)}\mathcal{A^+}, \mathcal{A^-}\mathcal{C}^{(2j)}\mathcal{A^-},
\]
where $\mathcal{C}^n$ means that state $\mathcal{C}$ is repeated $n$ times. Notice that any of the syllables leaves the system at $\mathcal{A^+}$ or $\mathcal{A^-}$ state and implies a certain change in the relative amount of pheromone $\D$ and hence, in the probabilities for the transition from one state to the other. Therefore, we can compute the probabilities of being left in state $\mathcal{A^+}$(resp. $\mathcal{A^-}$) with a certain value of $\D$ after the syllable $\ell+1$ as a function of the probabilities of each syllable to occur and the change in $\D$ that they produce. These can be easily computed, by induction, to be:
\begin{enumerate}
\item From state $\mathcal{A^+}$ to state $\mathcal{A^+}$: $p=\frac{(1+\e\D)^2}{4}$, $\D\mapsto \D+2$;
\item From state $\mathcal{A^-}$ to state $\mathcal{A^-}$: $p=\frac{(1+\e\D)^2}{4}$, $\D\mapsto \D+2$;
\item From state $\mathcal{A^+}$ to state $\mathcal{A^+}$ passing through state $\mathcal{B}$:
 $p=\frac{(1-\e\D)^2}{4}$, $\D\mapsto \D-2$;
\item From state $\mathcal{A^-}$ to state $\mathcal{A^-}$ passing through state $\mathcal{B}$:
 $p=\frac{(1-\e\D)^2}{4}$, $\D\mapsto \D-2$;
\item From state $\mathcal{A^-}$ to state $\mathcal{A^+}$ passing $n=2j+1$ times through state $\mathcal{C}$:
 $$p=\frac{1}{2^{2j+3}}+\e \frac{2j^2-4j}{2^{2j+3}}-\e \D \frac{2j-1}{2^{2j+3}}+\mathcal{O}(\e^2), j=0,1,\ldots,$$
 $$\D\mapsto \D-(2j-1), j=0,1,\ldots;$$
\item From state $\mathcal{A^+}$ to state $\mathcal{A^+}$ passing $n=2j$ times through state $\mathcal{C}$:
 $$p=\frac{1}{2^{2j+2}}+\e \frac{2j^2-6j+2}{2^{2j+2}}-\e \D \frac{2j-2}{2^{2j+2}}+\mathcal{O}(\e^2), j=1,2,\ldots,$$
 $$\D\mapsto \D-(2j-2), j=1,2,\ldots;$$
\item From state $\mathcal{A^+}$ to state $\mathcal{A^-}$ passing $n=2j+1$ times through state $\mathcal{C}$:
 $$p=\frac{1}{2^{2j+3}}+\e \frac{2j^2-4j+1}{2^{2j+3}}-\e \D \frac{2j-1}{2^{2j+3}}+\mathcal{O}(\e^2), j=0,1,\ldots,$$
 $$\D\mapsto \D-(2j-1), j=0,1,\ldots;$$
\item From state $\mathcal{A^-}$ to state $\mathcal{A^-}$ passing $n=2j$ times through state $\mathcal{C}$:
 $$p=\frac{1}{2^{2j+2}}+\e \frac{2j^2-6j+4}{2^{2j+2}}-\e \D \frac{2j-2}{2^{2j+2}}+\mathcal{O}(\e^2), j=1,2,\ldots,$$
 $$\D\mapsto \D-(2j-2), j=1,2,\ldots;$$
\end{enumerate}

If we set
\[
p(N+1,(\D,\mathcal{A^-}))=p_{N+1}(\D^-),
\]
\[
p(N+1,(\D,\mathcal{A^+}))=p_{N+1}(\D^+),
\]
for the probabilities of having a relative reinforcement $\D$ and end after the $N+1$ syllable at state $\mathcal{A^-}$ or $\mathcal{A^+}$ respectively, then one can easily write master equations for $p_{N+1}(\D^-)$ and $p_{N+1}(\D^+)$ using the probabilities of each syllable (points $1$ to $8$ above). If we perform the approximation
\[
p(\D^{+,-}+\delta)=p(\D^{+,-})+\delta \frac{\partial p}{\partial \D}(\D^{+,-}),
\]
in the master equations, then after straightforward calculations we arrive at the following equations for
$p_{N+1}(\D^-)$ and $p_{N+1}(\D^+)$:

\begin{equation}\label{probAneg2}
\begin{split}
p_{N+1}(\D^-)&=\frac 23 p_N(\D^-)+\frac 13 p_N(\D^+)-\frac{62}{27}\e p_N(\D^-) \\
&-\frac{10}{27}\e p_N(\D^+)-\langle \delta \D\rangle_{A^-\rightarrow A^-}\frac{\partial p_N}{\partial \D}(\D^-)\\
&- \langle \delta \D\rangle_{A^+\rightarrow A^-} \frac{\partial p_N}{\partial \D}(\D^+)+ \mathcal{O}(\D^2,\e^2),
\end{split}
\end{equation}
where
\[
\begin{split}
\langle \delta \D\rangle_{A^-\rightarrow A^-}&= 2\e\D-\frac 19 - \frac 49 \e+ \frac{44}{27}\e+2\e\D\sum_{j=1}^\infty \frac{(2j-2)^2}{2^{2j+2}}\\
&=-\frac 19 +\frac{32}{27}\e+\frac{64}{27}\e\D, \\
\langle \delta \D\rangle_{A^+\rightarrow A^-}&= \frac 19 - \frac 19 \e+ \frac{37}{27}\e+2\e\D\sum_{j=0}^\infty \frac{(2j-1)^2}{2^{2j+3}}\\
&=\frac 19 +\frac{34}{27}\e+\frac{17}{27}\e\D,
\end{split}
\]
and
\begin{equation}\label{probApos2}
\begin{split}
p_{N+1}(\D^+)&=\frac 13 p_N(\D^-)+\frac 23 p_N(\D^+)-\frac{19}{27}\e p_N(\D^-) \\
&-\frac{71}{27}\e p_N(\D^+)-\langle \delta \D\rangle_{A^-\rightarrow A^+}\frac{\partial p_N}{\partial \D}(\D^-)\\
&- \langle \delta \D\rangle_{A^+\rightarrow A^+}\frac{\partial p_N}{\partial \D}(\D^+)+ \mathcal{O}(\D^2,\e^2),
\end{split}
\end{equation}
where
\[
\begin{split}
\langle \delta \D\rangle_{A^-\rightarrow A^+}&= \frac 19 - \frac 29 \e+ \frac{37}{27}\e+2\e\D\sum_{j=0}^\infty \frac{(2j-1)^2}{2^{2j+3}}\\
&=\frac 19 +\frac{31}{27}\e+\frac{17}{27}\e\D,
\end{split}
\]
\[
\begin{split}
\langle \delta \D\rangle_{A^+\rightarrow A^+}&= 2\e\D- \frac 19 - \frac 29 \e+ \frac{44}{27}\e+2\e\D\sum_{j=1}^\infty \frac{(2j-2)^2}{2^{2j+2}}\\
&=-\frac 19 +\frac{38}{27}\e+\frac{64}{27}\e\D.
\end{split}
\]

Next we proceed to solve equations \eqref{probAneg2},\eqref{probApos2} both for $\e=0$ and $\e>0$.

%%%%%%%%%%%%%%%%%%%%%%%%%%%%%%%%%%%%%%%%%%%%%%%%%%%%%%%%%%%%
\emph{Equations at $\mathcal{O}(\varepsilon^0)$}
%%%%%%%%%%%%%%%%%%%%%%%%%%%%%%%%%%%%%%%%%%%%%%%%%%%%%%%%%%%%%%%

From (\ref{probAneg2}) and (\ref{probApos2}) we have that

\begin{equation}\label{eq1ordercero}
p_{N+1}(\D^-)=\frac 23 p_N(\D^-)+\frac 13 p_N(\D^+)+ \frac 19 \frac{\partial p_N}{\partial \D}(\D^-)- \frac 19 \frac{\partial p_N}{\partial \D}(\D^+),
\end{equation}

\begin{equation}\label{eq2ordercero}
p_{N+1}(\D^+)=\frac 13 p_N(\D^-)+\frac 23 p_N(\D^+)- \frac 19 \frac{\partial p_N}{\partial \D}(\D^-)+ \frac 19 \frac{\partial p_N}{\partial \D}(\D^+).
\end{equation}

By adding (\ref{eq1ordercero}) and (\ref{eq2ordercero}) we get
\begin{equation}\label{eq3ordercero}
p_{N+1}\equiv p_{N+1}(\D^-)+p_{N+1}(\D^+)=p_{N}(\D^-)+p_{N}(\D^+)\equiv p_{N}.
\end{equation}

Subtracting (\ref{eq2ordercero}) and (\ref{eq1ordercero}) we have
\begin{equation}\label{eq4ordercero}
\begin{split}
\delta p_{N+1}&\equiv p_{N+1}(\D^+)-p_{N+1}(\D^-)=-\frac 13 p_{N}(\D^-)+\frac 13 p_{N}(\D^+)\\
& -\frac 29 \frac{\partial p_N}{\partial \D}(\D^-)+
\frac 29\frac{\partial p_N}{\partial \D}(\D^+)\equiv \frac 13 \delta p_{N}+ \frac 29  \frac{\partial \delta p_N}{\partial \D}.
\end{split}
\end{equation}

Since
\[
\delta p_{N+1}-\delta p_{N}\simeq \frac {\partial \delta p}{\partial t},
\]
then (\ref{eq4ordercero}) becomes

\begin{equation}\label{eqordercero}
\frac {\partial \delta p}{\partial t}=-\frac 23 \delta p+ \frac 29 \frac {\partial \delta p}{\partial \D},
\end{equation}
and so
\begin{equation} \label{eq:solu}
\delta p= \exp(-\frac 23 t) \delta p_0 (\D+ \frac 29 t),
\end{equation}
where $\delta p_0$ is the initial data. This solution shows us an exponential decay for the difference of probabilities $p_N(\D^+)-p_N(\D^-)$ while the sum of probabilities remains constant for any $\D$ \eqref{eq3ordercero}. As a consequence, no preferential selection of any edge takes place. Directional persistence is not sufficient at this order for shortest path's selection. Next, we will discuss whether lower order terms are able to explain this preferential selection or one needs to invoke different effects.

%%%%%%%%%%%%%%%%%%%%%%%%%%%%%%%%%%%%%%%%%%%%%%%%%%%%%%%%%%%%
\emph{Equations at $\mathcal{O}(\varepsilon^1)$}
%%%%%%%%%%%%%%%%%%%%%%%%%%%%%%%%%%%%%%%%%%%%%%%%%%%%%%%%%%%%%%%

Subtracting $p_N(\D^-)$ from (\ref{probAneg2}) and $p_N(\D^+)$ from (\ref{probApos2}) we have
\begin{equation}\label{eq3orderone}
\begin{split}
\frac{\partial p^-}{\partial t}& =\frac 13 (p^+-p^-)-\frac 19 (p^+-p^-)_\D- p_\D^- \big(\frac {32}{27}\e+\frac{64}{27}\e\D\big) \\
& - p_\D^+ \big(\frac{34}{27} \e+\frac{17}{27}\e\D\big)-\frac{62}{27}\e p^--\frac{10}{27}\e p^+,
\end{split}
\end{equation}

\begin{equation}\label{eq4orderone}
\begin{split}
\frac{\partial p^+}{\partial t}& =\frac 13 (p^--p^+)-\frac 19 (p^--p^+)_\D- p_\D^- \big(\frac {31}{27}\e+\frac{17}{27}\e\D\big) \\
& - p_\D^+ \big(\frac{38}{27} \e+\frac{64}{27}\e\D\big)-\frac{19}{27}\e p^- -\frac{71}{27} \e p^+,
\end{split}
\end{equation}
where
\[
 p^+=p_{N}(\D^+),\qquad p^-=p_{N}(\D^-),
\]
\[
\frac{\partial p^-}{\partial t}= p_{N+1}(\D^-)-p_{N}(\D^-), \frac{\partial p^+}{\partial t}= p_{N+1}(\D^+)-p_{N}(\D^+),
\]
\[
(p^+)_\D=\frac{\partial p_N}{\partial \D}(\D^+),\qquad (p^-)_\D=\frac{\partial p_N}{\partial \D}(\D^-).
\]

Adding equations (\ref{eq3orderone}) and (\ref{eq4orderone}), putting $p=p^++p^-$ and making the approximations
$p^+=\frac p2+ \frac {\delta p}{2}, p^-=\frac p2- \frac {\delta p}{2}$ we have

\begin{equation}\label{eq5orderone}
\begin{split}
\frac{\partial p}{\partial t}& = -\frac 12 \big(5\e+6\e\D \big)p_\D - \frac 16 \e (\delta p)_\D- 3\e p\\
&=\Big((-\frac 52 \e -3 \e \Delta)p \Big)_{\Delta}- \frac 16 \e (\delta p)_\D,
\end{split}
\end{equation}
where we have also approximated the derivatives by
\[
p_\D^+=\frac 12 p_\D+\frac 12 (\delta p)_\D,\qquad p_\D^-=\frac 12 p_\D-\frac 12 (\delta p)_\D,
\]
and $\delta p$ is given by \eqref{eq:solu}.
Neglecting $\delta p$ which is exponentially decreasing in time (see equation \eqref{eq:solu}), equation \eqref{eq5orderone} becomes a transport equation that can be solved using
the characteristics method. The characteristics are the solution of
\[
\frac{d \D}{dt}=\frac 52 \e+3\e\D \Rightarrow \D= \frac 56 (\kappa \exp(3\e t)-1),
\]
and hence
\begin{equation}\label{solution}
\begin{split}
p&=\exp (-3\e t) p_0\big(\D-\frac 56(\exp(3\e t)-1)\big)\\
& \simeq \exp (-3\e t) p_0 (\D-\frac 56 3 \e t)= \exp (-3\e t) p_0 (\D-\frac 52\e t),
\end{split}
\end{equation}
since $\e t \ll 1$ and where $p_0$ is the initial data.

Notice that the probability distribution shifts towards increasing values of $\D$ at a velocity $\frac 52 \e$, \eqref{solution}. Hence, we conclude that shortest path (the one producing increase of $\D$) is progressively reinforced. This result provides an analytical proof support for the fact that both reinforcement and persistence are sufficient to produce shortest path selection at least for two ants. Remind that such shortest path selection was not possible with only one ant. In the next section we will consider the problem for larger number of ants.

%%%%%%%%%%%%%%%%%%%%%%%%%%%%%%%%%%%%%%%%%%%%%%%%%%%%%
\subsection{Non-reinforced network with $H$ ants}
%%%%%%%%%%%%%%%%%%%%%%%%%%%%%%%%%%%%%%%%%%%%%%%%%%%%%

Now, we consider the three node network in figure \ref{network} right but for $H$ ants.
Our analysis with two ants without reinforcement lead, from formulas \eqref{probAneg2} and
\eqref{probApos2} to a system that can be written in the form%
\[
\left(
\begin{array}{c}
P(\Delta ^{-}) \\
P(\Delta ^{+})%
\end{array}%
\right) _{N+1}=A\left(
\begin{array}{c}
P(\Delta ^{-}) \\
P(\Delta ^{+})%
\end{array}%
\right) _{N}+B\frac{\partial }{\partial \Delta }\left(
\begin{array}{c}
P(\Delta ^{-}) \\
P(\Delta ^{+})%
\end{array}%
\right) _{N},
\]%
where%
\[
A=\left(
\begin{array}{cc}
\frac{2}{3} & \frac{1}{3} \\
\frac{1}{3} & \frac{2}{3}%
\end{array}%
\right) ,\ B=\left(
\begin{array}{cc}
\frac{1}{9} & -\frac{1}{9} \\
-\frac{1}{9} & \frac{1}{9}%
\end{array}%
\right),
\]
and the states $\mathcal{A}^{-}$ and $\mathcal{A}^{+}$ correspond to $1$ or $2$
ants in the nest respectively. Notice that the elements of both the rows and
columns of matrix $A$ sum one and they are positive (since they correspond
to probabilities). Moreover, matrix $A$ is symmetric. These properties are
also verified when considering $H$ ants and, therefore, $H$ possible states
$\mathcal{A}^{1},\mathcal{A}^{2},\ldots,\mathcal{A}^{H}$ corresponding to $1,2,\ldots,H$
ants at the nest respectively. We can also write the system
\begin{equation}
\left(
\begin{array}{c}
P(\Delta ^{1}) \\
P(\Delta ^{2}) \\
\vdots  \\
P(\Delta ^{H})%
\end{array}%
\right) _{N+1}=A\left(
\begin{array}{c}
P(\Delta ^{1}) \\
P(\Delta ^{2}) \\
\vdots  \\
P(\Delta ^{H})%
\end{array}%
\right) _{N}+B\frac{\partial }{\partial \Delta }\left(
\begin{array}{c}
P(\Delta ^{1}) \\
P(\Delta ^{2}) \\
\vdots  \\
P(\Delta ^{H})%
\end{array}%
\right) _{N},  \label{system_states}
\end{equation}%
with $A$ a symmetric matrix with positive entries such that any row and
column sums one. This characterizes $A$ as an stochastic matrix which is,
moreover, symmetric. The Perron-Frobenius theorem implies then that there
exists an eigenvalue $\lambda =1$ and all other eigenvalues $\lambda$ are
such that $\left\vert \lambda \right\vert <1$. Since the matrix is
symmetric, such eigenvalues are real. The eigenvector corresponding to the
eigenvalue $1$, called the Perron-Frobenius eigenvector, is $(1,1,\ldots,1)^{T}$%
. This implies that $A=Q^{-1}DQ$ with $D$ the matrix of eigenvalues.
The lack of reinforcement implies that the probabilities $P(\Delta ^{1}), P(\Delta ^{2}), \ldots, P(\Delta ^{H})$
are indeed independent of $\Delta$ and therefore we will denote them as $P_1, P_2, \ldots, P_H$.
For all these reasons, \eqref{system_states} can be written in the form
\begin{equation}
\widetilde{P}_{N+1}=D\widetilde{P}_{N},  \label{sys_diag}
\end{equation}
where $\widetilde{P}_{N+1}=Q\left(P_1, P_2,\ldots,P_H\right) _{N+1}^{T}$.

Notice that (\ref{sys_diag}) is the discretized version of the following system of equations
\begin{equation}
\frac{\partial \widetilde{P}}{\partial t}=(D-I)\widetilde{P},
\label{trans_r}
\end{equation}
with solution $\widetilde{P}(t)=\exp((D-I)t) \widetilde{P}_{0}$.
Notice also that all components of $D-I$, except for the first one (corresponding to the
Perron-Frobenius eigenvalue $\lambda=1$ for $A$) are negative and hence
\[
e^{(D-I)t}=\left(
\begin{array}{cccc}
1 & 0 & \cdots  & 0 \\
0 & e^{ (\lambda _{2}-1)t} & 0 & \vdots  \\
\vdots  & 0 & \ddots  & 0 \\
0 & \cdots  & 0 & e^{(\lambda _{H}-1)t}%
\end{array}%
\right),
\]%
with $\lambda _{j}<1$, $j=2,\ldots,H$. The eigenvectors $e_{j}$
corresponding to eigenvalues $\lambda _{j}$ , $j=2,...,H$, are orthogonal to
the Perron-Frobenius eigenvector. Hence the vectors with all entries equal
to $0$ except for a $1$ at position $k$ and a $-1$ at position $l$, which are
orthogonal to the vector $(1,1,...,1)^{T}$, are linear combinations of the
eigenvectors $e_{j}$, $j>1$. This implies that
\begin{equation}
\left\vert P_{k,N}-P_{l,N}\right\vert \leq
Ce^{-\inf_{j>1}\left\{ \left\vert  (\lambda _{j}-1)\right\vert \right\} t}%
\text{ \ for all }k,l.  \label{expo_conv}
\end{equation}%
Since the probability distribution converge to equilibrium exponentially fast,
we will not make distinction among different states in what
follows and we will merely write equations for%
\[
p(N)=\sum_{j=1}^{H}P_{j}
\]
or, in other words, for the component of the probability vector $\left(
P_1,P_2,\ldots,P_H\right) _{N+1}^{T}$ on the
Perron-Frobenius eigenstate (all the other components converge exponentially
fast to zero by (\ref{expo_conv})).

%%%%%%%%%%%%%%%%%%%%%%%%%%%%%%%%%%%%%%%%%%%%%%%%%%%%%
\subsection{Reinforced network with $H$ ants}
%%%%%%%%%%%%%%%%%%%%%%%%%%%%%%%%%%%%%%%%%%%%%%%%%%%%%
Finally, we consider the three node network in figure \ref{network} right with reinforcement and for $H$ ants.
As in the case for two ants, we can decompose the evolution as a sequence of syllables starting and ending in a state $\mathcal{A}$ at which all ants are at the nest or the food source. All the possible syllables are of the form $\mathcal{A} \mathcal{A}$, $\mathcal{A}\mathcal{B}\mathcal{A}$ and $\mathcal{A}\mathcal{C}^(j)\mathcal{A}$, where the state $\mathcal{B}$ consists of all ants in the node that is not nest nor food source and state $\mathcal{C}$ can be any possible combination of $n_1$ ants at the nest or food source and $n_2$ ants at the other vertex. Since there are arbitrary long sequences of $\mathcal{C}$ states we will denote by $n_1^(1), n_1^(2), \ldots$ the number of ants that are at the nest or the food source at each of the $\mathcal{C}$ states of the sequence. Similarly, we denote by $n_2^(1), n_2^(2), \ldots$ the number of ants that are not at the nest nor at the food source. Notice then that $n_1^(j)+n_2^(j)=H$. Following the same steps as in the case for two ants, we can then write the following equation for the probability:
\begin{equation}\label{ecu:H2}
\begin{split}
& p(\D,N+1)\\
&=\quad\frac{(1+\e\D)^H}{2^H}p(\D-H,N)+\frac{(1-\e\D)^H}{2^H}p(\D+H,N)\\
&\quad +\sum \frac{1}{2^{(S+1)H-x}}\left( \prod_{j=1}^{S+1}\binom{H-n_2^{(j-1)}}{H-(n_2^{(j)}+n_2^{(j-1)})} \right)\\
&\quad \quad \left(1+\e \mathcal{Q}_{\{S,n_2^{(1)},\ldots,n_2^{(S)}\}}^H(\D)\right)p(\D-((S+1)H-3x),N),
\end{split}
\end{equation}
where the sum goes over all $S$ from $1$ to $\infty$ and over all the $S-$uplas $\left\{n_2^{(1)},n_2^{(2)},\ldots,n_2^{(S)}\right\}$
such that $1\leq n_2^{(k)}< H$, $n_2^{(k)}+n_2^{(k+1)}\leq H, \forall k\geq 1$, $x=\sum_{k=1}^S n_2^{(k)}$ and
\begin{equation}\label{ecu:H3}
\begin{split}
&\mathcal{Q}_{\{S,n_2^{(1)},\ldots,n_2^{(S)}\}}^H(\D)\\
&\quad =(n_1^{(1)}-n_2^{(1)})(\D-(S+1)H+3x)\\
&\quad +\sum_{j=1}^{S}\left((n_1^{(j)}-2n_2^{(j+1)})\Big(\sum_{k=1}^{j}(n_1^{(k-1)}-n_2^{(k)})\right.\\ %HAY QUE CERRAR LOS PARENTESIS
&\quad \left. -\frac 12 \left(\sum_{k=1}^{j}n_2^{(k)}+n_2^{(k-1)}\right)+(\D-(S+1)H+3x) \Big) \right).
\end{split}
\end{equation}

Applying Taylor's expansion in \eqref{ecu:H2} and using the relations
\begin{equation}\label{num1}
\frac{2}{2^H}+\sum \frac{1}{2^{(S+1)H-x}}\left(\prod_{j=1}^{S+1}\binom{H-n_2^{(j-1)}}{H-(n_2^{(j)}+n_2^{(j-1)})} \right)=1,
\end{equation}
\begin{equation}\label{num2}
\sum \frac{(S+1)H-3x}{2^{(S+1)H-x}}\left(\prod_{j=1}^{S+1}\binom{H-n_2^{(j-1)}}{H-(n_2^{(j)}+n_2^{(j-1)})} \right)=0,
\end{equation}
(see remark at the end of this chapter for a proof of this formulas) we arrive, keeping up to $\mathcal{O}(\e)$ terms, at the equation

\begin{equation} \label{eq:c}
p(\D,N+1)=p(\D,N)+c(\D)\e\frac{\partial p}{\partial \D}(\D,N),
\end{equation}
where

\begin{equation} \label{velocidad}
c(\D)=-\sum \frac{(S+1)H-3x}{2^{(S+1)H-x}}\left(\prod_{j=1}^{S+1}\binom{H-n_2^{(j-1)}}{H-(n_2^{(j)}+n_2^{(j-1)})} \right)\mathcal{Q}_{\{S,n_2^{(1)},\ldots,n_2^{(S)}\}}^H(\D).
\end{equation}

Equation \eqref{eq:c} is a discretized version of the transport equation
\[
\frac{\partial p(\Delta,t)}{\partial t}=c \e \frac{\partial p}{\partial \Delta}(\Delta, t).
\]
The constant $c(0)$ is computed numerically from formulas \eqref{ecu:H2} and \eqref{ecu:H3}. Since we are considering the early times when the maximum of the probability distribution is close to $\D=0$, it is the convection produced by $c(0)$ what breaks the symmetry and shifts the probability distribution towards increasing/decreasing values of $\D$ at a velocity $c(0)$ provided it is strictly positive/negative. Our numerical computations yield the values for the velocity $c(0)=11.2476$ for $3$ ants and $c(0)=28.2320$ for $4$ ants. Hence, the shortest path will be progressively reinforced and this will occur at a velocity that increases with the number of ants.

%%%%%%%%%%%%%%%%%%%%%%%%%%%%%%%%%%%%%%%
\subsubsection{The limit $\e=0$}
%%%%%%%%%%%%%%%%%%%%%%%%%%%%%%%%%%%%%
By considering $\e=0$ in \eqref{ecu:H2} we return to the problem without reinforcement and the equation for the probability is:
\begin{equation}\label{ecu:H}
\begin{split}
&p(\D,N+1)=\frac{1}{2^H}p(\D-H,N)+\frac{1}{2^H}p(\D+H,N)\\
&+\sum \frac{1}{2^{(S+1)H-x}}\left( \prod_{j=1}^{S+1}\binom{H-n_2^{(j-1)}}{H-(n_2^{(j)}+n_2^{(j-1)})} \right)p(\D-((S+1)H-3x),N)
\end{split}
\end{equation}
where the sum goes over all $S$ from $1$ to $\infty$ and over all the $S-$uplas $\left\{n_2^{(1)},n_2^{(2)},\ldots,n_2^{(S)}\right\}$
such that $1\leq n_2^{(k)}< H$ and $n_2^{(k)}+n_2^{(k+1)}\leq H, \forall k\geq 1$ and $x=\sum_{k=1}^S n_2^{(k)}$.
We suppose that $n_2^{(0)}=n_2^{(S+1)}=0$.

Applying Taylor's expansion in (\ref{ecu:H}) we have that

\begin{equation}\label{ecu2H}
\begin{split}
&p(\D,N+1)=\frac{1}{2^H}p(\D,N)+\frac{1}{2^H}p(\D,N)\\
&+\sum \frac{1}{2^{(S+1)H-x}}\left(\prod_{j=1}^{S+1}\binom{H-n_2^{(j-1)}}{H-(n_2^{(j)}+n_2^{(j-1)})}\right)p(\D,N)\\
&-\left[ \sum \frac{(S+1)H-3x}{2^{(S+1)H-x}}\left(\prod_{j=1}^{S+1}\binom{H-n_2^{(j-1)}}{H-(n_2^{(j)}+n_2^{(j-1)})} \right)\right]\frac{\partial p}{\partial \D}(\D,N)+\mathcal{O}(\D^2)
\end{split}
\end{equation}

As a final remark, we note that without reinforcement the convection velocity vanishes by formula \eqref{num2}, $p(\D,N+1)=p(\D,N)$ by formula \eqref{num1}, and hence no path selection takes place. Formula \eqref{num1} follows from the fact that the sum of all the probabilities must to be one and formula \eqref{num2} must necessarily be true due to the following:
\begin{enumerate}
  \item If there is not reinforcement, the movement of just one ant does not affect the others.
  \item If the initial position for all ants is at node $\mathrm{1}$, we have with probability $1$ that all the ants will be at node $\mathrm{1}$ at a subsequent time $t$.
  \item The mean $\D$ change produced by an ant with initial position at node $\mathrm{1}$ and coming for the first time also to node $\mathrm{1}$ is
      \[
      \begin{split}
        \langle \delta \D\rangle&=2p_{\mathrm{1},\mathrm{2},\mathrm{1}}+(-2)p_{\mathrm{1},\mathrm{3},\mathrm{2},\mathrm{3},\mathrm{1}}+
        0p_{\mathrm{1},\mathrm{3},\mathrm{2},\mathrm{1}}+0p_{\mathrm{1},\mathrm{2},\mathrm{3},\mathrm{1}}\\
        &\qquad =2\frac 14+(-2)\frac 14+0+0=0.
        \end{split}
      \]
\end{enumerate}
When all the ants meet at node $\mathrm{1}$, each ant has done a certain number of elementary paths: $(\mathrm{1},\mathrm{2},\mathrm{1}),
(\mathrm{1},\mathrm{3},\mathrm{2},\mathrm{3},\mathrm{1}), (\mathrm{1},\mathrm{3},\mathrm{2},\mathrm{1}), (\mathrm{1},\mathrm{2},\mathrm{3},\mathrm{1})$. Since the mean change $\langle \delta \D\rangle=0$, then \eqref{num2} must be true.

The computational time grows exponentially with $H$ so that we can only compute a few values of summands in \eqref{velocidad} (for $\D =0$) when $H<5$. Nevertheless, the tendency to reach larger values of $c(0)$ can be clearly appreciated at least for $H=5,6,7$.

%%%%%%%%%%%%%%%%%%%%%%%%%%%%%%%%%%%%%%%%%%%%%%%%%%%%%%%%%%%%%%%%%%%%%%
\section{Conclusion}
%%%%%%%%%%%%%%%%%%%%%%%%%%%%%%%%%%%%%%%%%%%%%%%%%%%%%%%%%%%%%%%%%%%%%%

We have presented a model for ants to simulate their behavior when foraging.
It is well know that social insects, as for example ants, leave a trail
to coordinate the group and to communicate to each other. This pheromone
plays an important role to recruit the individuals and reinforce the shortest
path between nest and food source. We have shown by means of numerical
simulations and analytical arguments that in order for the ants to follow the geodesic path
in a two or three node network, it is necessary
not only to invoke the pheromone-induced reinforcement but also to have a directionality constraint.
Such constraint is so that ants prefer to maintain their direction of motion to turn back and return.
Furthermore, more than one ant is also needed to reinforce the geodesic path, with the velocity of reinforcement
increasing exponentially fast with the number of ants.
We expect that the combined effect of reinforcement and persistence is able to induce the formation of ant trails of minimal length not only in simple networks, but also in more complex networks, in the plane or in surfaces of general topology. This is the object of our current research and results will be presented in future publications.

%%%%%%%%%%%%%%%%%%%%%%%%%%%%%%%%%%%%%%%%%%%%%%%%%%%%%%%%%%%%%%%%%%%%%%
\textbf{Acknowledgements}
%%%%%%%%%%%%%%%%%%%%%%%%%%%%%%%%%%%%%%%%%%%%%%%%%%%%%%%%%%%%%%%%%%%%%%
This work has been supported by the Spanish Ministry of research through projects
$MTM2008-03255$ and $MTM2007-61755$.

%%%%%%%%%%%%%%%%%%%%%%%%%%%%%%%%%%%%%%%%%%%%%%%%%%%%%%%%%%%%%%%%%%%%%%

%%%%%%%%%%%%%%%%%%%%%%%%%%%%%%%%%%%%%%%%%%%%%%%%%%%%%%%%%%%%%%%%%%%%%%%

%%%%%%%%%%%%%%%%%%%%%%%%%%%%%%%%%%%%%%%%%%%%%%%%%%%%%%%%%%%%%%%%%%%%%%%
\end{document}